\documentclass[11pt]{article}

\usepackage[usenames,dvipsnames,table]{xcolor}
\usepackage[utf8]{inputenc}

\pdfoutput=1 % if your are submitting a pdflatex (i.e. if you have
\usepackage{jheppub} % for details on the use of the package, please
\usepackage{graphicx}
\usepackage{amsmath, amssymb}
\usepackage{comment}

\usepackage{mathrsfs}

\usepackage{framed}
\definecolor{shadecolor}{rgb}{0.90,0.90,0.90}

\usepackage{mathtools}

\def\be{\begin{eqnarray}}
\def\ee{\end{eqnarray}}
\def\bea{\begin{eqnarray}}
\def\eea{\end{eqnarray}}

\newcommand{\Tr}{\mbox{Tr}}

\newcommand{\beal}{\begin{equation}
\begin{aligned}}
\newcommand{\eeal}{\end{aligned}
\end{equation}}
\newcommand{\bem}{\begin{multline}}
\newcommand{\eem}{\end{multline}}

\def\U{\text{U}}
\def\SU{\text{SU}}

\def\SO{\text{SO}}

\begin{document}

\title{Global symmetries: locality, unitarity, and regularity}

\medskip

\author{Ibrahima~Bah,$^{1}$  Shlomo~S.~Razamat,$^{2}$ Michal Shemesh,$^{2}$ and Hannah~Tillim$^{1,3}$}
 %\email{iboubah@jhu.edu, razamat@physics.technion.ac.il, }

 \medskip

\affiliation{${}^{1}$ William H. Miller III Department of Physics and Astronomy, Johns Hopkins University,\\
${}\;\;$ 3400 North Charles Street, Baltimore, MD 21218, USA}

\affiliation{${}^{2}$ Department of Physics, Technion, Haifa, 32000, Israel}

\affiliation{${}^{3}$ Capital Fund Management, 23 Rue de l'Universit\'e, 75007 Paris, France}

%%%%%%%%%%%%%%%%%%%%%%%%%%%%%%%%%%%%%%%%%%%%%%%%%%%%%%%%%%%%%%%%%%%%%%%%%%%%%%%%%%%%%%%%%%%%%%%%%%%%%%%%%%%%%%%%%%%%%%%%
%%%%%%%%%%%%%%%%%%%%%%%%%%%%%%%%%%%%%%%%%%%%%%%%%%% ABSTRACT %%%%%%%%%%%%%%%%%%%%%%%%%%%%%%%%%%%%%%%%%%%%%%%%%%%%%%%%%%%
%%%%%%%%%%%%%%%%%%%%%%%%%%%%%%%%%%%%%%%%%%%%%%%%%%%%%%%%%%%%%%%%%%%%%%%%%%%%%%%%%%%%%%%%%%%%%%%%%%%%%%%%%%%%%%%%%%%%%%%%

\abstract{
We study the apparent tension between locality and unitarity for symmetries in quantum field theory.  This emerges in the context of categorical symmetries where symmetry operators are generically non-invertible.  We argue that locality imposes particular regularities in the action of symmetries on the Hilbert space.  This allows us to introduce an observable that can measure the properties of the non-locality for symmetry operators.  We study it for a  class of examples and demonstrate that this observable can encode data associated to the fusion algebra of symmetries.}

\maketitle

\section{Introduction}

The study of symmetry in quantum systems has taken center stage in physics and mathematics in recent times.  This is mainly due to the realization that symmetries, and their various generalizations and structures, are characterized by the properties of topological operators in quantum systems \cite{Gaiotto:2014kfa,Tachikawa:2017gyf}, {\it i.e.} operators that are preserved under small deformations of spacetime.

In Quantum Mechanics (QM), QFT in $0+1$ dimensions, the classification of symmetries was performed by Wigner, with the conclusion that these are given by (anti)unitary operators $U_\alpha$ commuting with the Hamiltonian.\footnote{To be more precise Wigner has classified operator which commute with the Hamiltonian {\it and} preserve probabilities. }  Since the Hamiltonian generates small deformations in time, this implies a topological nature for $U_\alpha$ along the time direction.  
On the other hand, the fact that the symmetry operators are (anti)unitary implies that they form the mathematical structure of {\it a group} ($G$). In particular, every symmetry operator has an inverse. The Hilbert space and operators of a  given QM model decompose into irreducible representations of the symmetry group. Moreover, the spectrum of irreducible representations of the symmetry group $G$ realized on the Hilbert space of a given QM system is {\it arbitrary}.\footnote{For example, consider an $G=\SO(3)$ global symmetry. We can construct a Hilbert space of QM as a tensor product of an arbitrary collection and multiplicity of various spins and define a Hamiltonian acting on it preserving the $G=\SO(3)$ symmetry group.}

Let us turn to QFTs in $d+1$ dimensions. For concreteness, consider a QFT
which can be constructed as a thermodynamic/continuum limit of a QM system. 
When viewing the system thus  before taking the limit as a quantum mechanical model, Wigner's theorem still implies that the symmetries correspond to (anti)unitary operators, which build the structure of a group. 
However, QFTs in higher dimensions can enjoy the property of {\it locality}.
This imposition significantly constrains the Hamiltonian of the initial QM system. 
The unitary operators commuting with the Hamiltonian may not respect this notion of locality. 
In particular, representations of operators might mix local operators acting on the Hilbert space with non-local ones.
The standard treatment of group-like symmetries in a QFT restricts to symmetries which respect {\it locality}, {\it i.e.} representations which take local operators to local operators. This treatment was in recent years extended to considering symmetry operators which commute with the Hamiltonian and  which respect some version of the notion of locality, but do not necessarily obey the mathematical structure of a group. 

More precisely, following \cite{Gaiotto:2014kfa}, one can think of a sub-class of operators which commute with the Hamiltonian
and which are {\it topological}. In particular, operators ${\cal O}_{\cal M}$ in this sub-class can be associated to co-dimension $n$ surfaces, ${\cal M}$, in space-time. The correlation functions of  ${\cal O}_{\cal M}$ depend on the choice of ${\cal M}$ only topologically. This generalizes the independence of $U_\alpha$ with respect to time in QM (up to crossing charged operators).  Its topological nature implies that we are free to deform the operator and non-trivial phenomena occur only when it touches other insertions.  The action of the operators is purely dictated by Topological Field Theory (TQFT) rules \cite{Atiyah:1989vu} where expectation values can be obtained by arbitrary cutting and gluing of the spacetime.  In this sense topological operators are intimately related to the locality property of QFTs, {\it i.e.} they are synonymous with locally-acting symmetries.   
A QFT then might possess symmetry operators which are topological but either may not have an inverse at all ({\it e.g.} involve projectors on the Hilbert space), or possess an inverse which itself  does not act locally and thus is not a topological operator.\footnote{Example of the former is the duality defect of the $2d$ Ising model, while examples of the latter can be found in $2d$ CFT examples having topological Verlinde lines realizing the Fibonacci category, {\it e.g} the $({\frak g}_2)_1$ WZW model. See for example \cite{Lin:2023uvm}. } Such symmetries are often called {\it non-invertible}. See \cite{Shao:2023gho,Schafer-Nameki:2023jdn,Bhardwaj:2023kri} for reviews. Thus, inherently,  non-invertibility of symmetries is a tension between locality and unitarity.

The imposition of locality, and therefore topology, for symmetries has far-reaching mathematical implications. For example, the set of such operators  builds a structure of a {\it (higher) fusion category} \cite{douglas2018f,Dcoppet2023,decoppet2024,ferrer2024daggerncategories,JohnsonFreyd2022,bah2025physicshighercondensationdefects}. 
In recent years it has been understood that this mathematical structure is very natural in the study of QFTs: {\it e.g.} gauging symmetry groups leads in general to  theories with categorical symmetries \cite{Tachikawa:2017gyf}.\footnote{
We stress that not all the known non-invertible symmetries are of this sort \cite{Sun:2023xxv}.
} More fundamentally, categories are present even in the context of group ($G$)-like symmetries as the set of irreducible representations of a group has the structure of a fusion category, $\text{Rep}(G)$.

 Let us consider operators ${\cal O}_{\cal M}$ which are labeled by co-dimension one surfaces ${\cal M}$ in space time: the zero-form symmetries \cite{Gaiotto:2014kfa}. 
We can choose such operators to be located at a single point in time and spanning the spatial manifold: in this case we can refer to these as symmetry operators acting on the Hilbert space. Alternatively, we can consider them to be spanning the temporal direction and a co-dimension one surface in space. In such a case we can think of ${\cal O}_{\cal M}$ as modifying  the Hamiltonian of the system: these are symmetry defects.  The topological symmetry operators define the structure of a fusion algebra, while the topological defects promote this structure to that of a fusion category.
The structure of a fusion category has many important physical implication. See {\it e.g.} \cite{Chang:2018iay,Komargodski:2020mxz,vanBeest:2023dbu}.  We can refer to considering this categorical structure of symmetries as {\it a local approach}.
 
One can also consider a complementary view: instead of focusing on the categorical aspects of symmetries, we can  discuss unitary operators commuting with the Hamiltonian. By Wigner's theorem this is a set of natural operators in QM: if we find a non-unitary operator which commutes with the Hamiltonian, we could treat  it as a charge in QM. For example, such a charge can be always exponentiated to produce a unitary operator. As we consider here $d+1$ dimensional QFTs which are limits of QM systems, given a non-invertible symmetry one can always construct unitary operators from it. We stress again that these unitary operators are not in general consistent with TQFT rules in $d+1$ dimensions.
%They are however topological in the QM, one dimensional, perspective of the model. 
The set of {\it all} such operators forms the structure of a group which encodes in it  the non-invertible symmetries. We can refer to considering this group theoretic structure of symmetries as {\it a unitary approach}.

In this paper we will be interested in a relation between the two approaches centered on representations of the Hilbert space under the two notions of symmetry. For QM the representations of the Hilbert space under the symmetry is not constrained at all. However, it was argued that, under some assumptions, locality of the QFT does impose restrictions on the asymptotic decomposition of its Hilbert space into representations of the symmetry in the local approach. Focusing on zero-form symmetries, we will review this by illustrating the restriction in a variety of simple examples: the Hilbert space in fact forms a {\it regular} representation, properly defined.  
Certain deviations  from regularity in the unitary approach can be viewed as indicators  of non-locality.
They can also be viewed as a signal for non-invertible categorical symmetries by a basis transformation from unitary operators to topological ones.  

\

The paper is organized as follows. In  Section \ref{sec:regularity}
we discuss the relation between the locality properties of a QFT and the regularity of the decomposition of the Hilbert space into representations of the symmetry. We first discuss this for group-like symmetries and then for categorical ones. In Section \ref{sec:unitaryapproach} we discuss in detail the unitary approach. We focus  on QFTs whose categorical expressions have a finite number of simple objects. In particular we discuss a measure of irregularity of the spectrum. In Section \ref{sec:examples} we consider several examples of relations between local and unitary approaches to symmetries. Finally, in Section \ref{sec:discussion} we discuss our results.

\

\section{Local symmetries and regular representations}\label{sec:regularity}

In this section we explore the implication of locality  on symmetry operators in quantum field theory.\footnote{This should not be confused with local symmetry transformations which are used to define gauge symmetry.}  We will be interested in QFTs with a global symmetry $\mathcal{S}$, which may be a group or a fusion category.  We will restrict to $0$-form symmetries which act on local operators.  We also assume a choice of time direction and consider QFTs on $\mathbb{R}_t \times \mathcal{M}$, where $\mathcal{M}$ is a spatial slice.  For a given QFT, there is a Hilbert space of states, $\mathcal{H}$, defined on $\mathcal{M}$.  

Consider a symmetry operator $g_\alpha$ with an action $U^\alpha$ on $\mathcal{H}$. Locality, in its strongest form, means that for a disjoint union of two subregions $A$ and $B$ on $\mathcal{M}$, the Hilbert space on the union is a tensor product $\mathcal{H}_A \otimes \mathcal{H}_B$, and the action of $g_\alpha$ decomposes as $U^\alpha_A \otimes U^\alpha_B$. Here $U^\alpha_{A(B)}$ is a representation of $g_\alpha$ on $\mathcal{H}_{A(B)}$.\footnote{This perspective on locality is related to the splitability property discussed in \cite{BuchholzDoplicherLongo1986}.}

There are important caveats to the factorization of the Hilbert space, as stated.  For example, in a gauge theory, if we pick a subregion $A$, one needs to impose Gauss' law on the boundary, thus the Hilbert space on $A$ comes with a choice of a charge (representation) sector.  For every such choice, we can expect factorization and the Hilbert space of the full system is a sum over all possible choices.  Similarly, for theories with fermions, we will need to make a choice of spin structure and sum over such choices subject to some global constraints.  Even in these cases, where factorization is subject to global choices, we consider locality of $U^\alpha$ to mean a decomposition of its action as $U^\alpha_A \otimes U^\alpha_B$.  

Our primary goal in this section is to study the implications of the above picture of locality on the trace of $U^\alpha$ over the Hilbert space $\mathcal{H}$.  If we consider a finite-dimensional Hilbert space, then $\Tr_\mathcal{H}(U^\alpha)$ is well-defined.  In a general QFT, the trace need not exist. On the other hand, we can consider the QFT in Euclidean space with compact imaginary time of period $\beta$, and define the observable as the limit of the thermal trace
\begin{equation}
    C(\alpha) \equiv  \lim_{\beta \to 0} \frac{\Tr_\mathcal{H} \left(U^\alpha e^{-\beta H}\right)}{\Tr_\mathcal{H}\left( e^{-\beta H}\right)} \,,\label{eq:defCa}
\end{equation} where $H$ is the Hamiltonian of the system, and $\beta$ is the inverse temperature.  

\subsection{Invertible symmetries and regularity}

To understand $C(\alpha)$ in \eqref{eq:defCa}, let us consider a set-up where the strongest version of the factorization and locality condition holds.  Take a QFT on $\mathbb{R}_t \times {\cal M}$ for compact ${\cal M}$ with volume $V$.  We also assume that the QFT has a finite correlation length.  Now consider a decomposition of ${\cal M}$ into $N$ disjoint cells whose characteristic size is larger than the correlation length of the QFT.  We also restrict to a symmetry group $G$, with $g_\alpha \in G$, and the action the symmetry given by a representation $U^\alpha = \rho(g_\alpha)$.  

The factorization assumption is
\begin{equation}
\mathcal{H} = \bigotimes_i^N \mathcal{H}_i, \qquad \rho(g_\alpha) = \bigotimes_i^N \rho_{R_i}(g_\alpha),
\end{equation} where $i$ labels the cells and $R_i$ labels the representation of the Hilbert space in a cell. We assume that generically $R_i$ is not trivial. In this setup, the Hilbert spaces on each cell and the system Hilbert space are all finite dimensional for a fixed $N$. This picture is consistent with examples of field theories obtained from a lattice with on-site degrees of freedom, and with scalar field theories in a box of size $V$.  Since the various Hilbert spaces are finite dimensional, we can immediately take the limit $\beta\to 0$ and obtain
\begin{equation}
    C(\alpha) = \prod_i^N \frac{\chi_{R_i}(g_\alpha)}{\chi_{R_i}(e)}\,,
\end{equation} where $\chi_R(g_\alpha)$ is the character of the $R$ representation of $G$, and $e$ is the unit element.  Without loss of generality, we take each cell to be in the same representation $R$ and obtain, 
\begin{equation}
C(\alpha) = \left(\frac{\chi_R(g_\alpha)}{d_R} \right)^N \, \xrightarrow{N \to \infty } \, \delta_{g_\alpha, e}, \qquad \mbox{since} \qquad |\chi_R(g_\alpha)| \leq d_R
\end{equation} where $d_R = \chi_R(e)$ is the dimension of the representation.  In the limit of large $N$ we use the fact that for non-trivial elements $g_\alpha$, and representation $R$, the norm of the character is strictly bounded above by $d_R$ with equality only for $g_\alpha =e$. 

Let us consider a simple $U(1)$ example where ${\cal H}_i$ are two-dimensional, decomposing into $\pm q$ charged states. 
Then,
\be
\frac{\chi_R(g_\alpha)}{\chi_R(e)}=\frac{e^{i\, q\,\alpha}+e^{-i\, q\,\alpha}}2=\cos \,q\,\alpha\,,
\ee and thus if $\alpha\neq 0$ this is smaller than one in absolute value.
The intuitive picture here is that for any non-trivial representation an increasing tensor power will decompose into all possible irreducible representations with relative weights consistent with regular representations. For example, if the symmetry is $\U(1)$ the tensor product will produce all the possible charges with the same multiplicity in the large $N$ limit.\footnote{Note that we could have chosen in the example the two-dimensional on-site Hilbert space to be spanned by say two positive charges. Then naively only positive charges would have been generated. However, this situation is equivalent, by redefining symmetry generators with a phase, to having both negative and positive charges. This would have generated all the positive charges of $\U(1)/\Gamma$ with some choice of a  cyclic group $\Gamma$.}  We analyze this by decomposing $C(\alpha)$ into the characters of the irreducible representations of the group $G$ as
\begin{equation}
    C(\alpha) = \frac{1}{d_R^N} \sum_{a}\, m_a \,\chi_a (g_\alpha), \qquad m_a = \frac{1}{|G|} \sum_{g_\alpha \in G} \bar{\chi}_a(g_\alpha) \chi_R^N(g_\alpha)
\end{equation} where $a$ labels irreducible representations of $G$, $m_a$ is the multiplicity of the representation, and $|G|$ is the order of the group.  In writing the sum we are assuming discrete groups, for Lie groups we replace the sum with the Haar measure.  Now we wish to understand the large N limit of the multiplicity
\begin{equation}
    m_a =\frac{d_R^N}{|G|} \left( d_a+ \sum_{g_\alpha \neq e} \bar{\chi}_a(g_\alpha) \left(\frac{\chi_R(g_\alpha)}{d_R} \right)^N \right) \, \xrightarrow{N \to \infty } \, \frac{d_R^N}{|G|}\, d_a.  
\end{equation}  In evaluating the sum, we peeled off the identity element to obtain the leading term with $d_a = \chi_a(e)$.  Now we can write our observable as
\begin{equation}\label{eq:Cwithcharacters}
    C(\alpha) = \frac{1}{|G|}\sum_{a}\, d_a \,\chi_a (g_\alpha).
\end{equation}  We observe that $C(\alpha)$ vanishes for $g_\alpha\neq e$  because it is the character of the \textit{regular representation} of $G$.  This yields an important result: \textit{The Hilbert space $H$ of a local QFT in the limit of large energies is asymptotically dominated by the regular representation of locally-acting global symmetries.} The mathematically precise statement regards the quantity $C(\alpha)$. For this statement to make sense it is important that in the definition of $C(\alpha)$ \eqref{eq:defCa} one normalizes with $\Tr \,e^{-\beta\, H}$. The claim of {\it regularity} of representations should be understood as this asymptotic statement.
Note that the asymptotic behavior of $C(\alpha)$ was discussed before in various physical contexts. See {\it e.g.}  \cite{Harlow:2021trr,Lin:2022dhv,Benjamin:2023qsc,Pal:2020wwd}.\footnote{In particular, proofs of regularity properties conjectured in \cite{Harlow:2021trr} were discussed  in the framework of algebraic QFT both for finite and categorical symmetries \cite{Magan:2021myk,Benedetti:2024dku}.}

In evaluating these formulae we assume that the representation $R$ is non-trivial.  If $R$ is not faithful, then we should evaluate the sum over the $G/H$ where $H$ is a normal subgroup corresponding to elements that act trivially for the representation R.  Such a sum will lead to the same conclusion, and the Hilbert space would be in the regular representation of $G/H$.

To further establish the regularity of locally acting symmetries, we consider several non-trivial examples and compute $C(\alpha)$.  

\subsubsection*{ Lattice example} Let us consider a more concrete computation with a lattice system.
Consider  a circular lattice with a two-level system at each lattice site. On each site there is an $\SO(3)$ symmetry acting. Let us assume that the Hamiltonian is such that the diagonal $\SO(3)$ subgroup is preserved. An example of such a Hamiltonian is the Heisenberg chain. If the number $N$ of sites is finite  then the spectrum of representations is not regular. The spectrum is given by the decomposition of a tensor product of $N$ spin-half irreps. A direct computation shows that  spin $k$ appears with multiplicity $n_N^k$,
\be
n_N^k=\left(\begin{matrix}
N-1 \\
\frac{N+2k-2}2
\end{matrix}\right)-\left(\begin{matrix}
N-1 \\
\frac{N+2k+2}2
\end{matrix}\right) = N! \frac{2k+1}{\left( \frac{N}{2}+k+1 \right)!\left( \frac{N}{2}-k \right)!} \,.
\ee Thus, if $N$ is even, only integer spins appear and if $N$
is odd, only half-integer spin appears. In the former case the group is $\SO(3)$ with non-projective irreps and in the latter case it is $\SO(3)$ with only projective irreps. We observe that different representations appear with different multiplicities for fixed $N$. However, fixing two spins, $k$ and $k'$, and taking $N\to \infty$,
\be
\lim_{N\to \infty}\;\frac{n_N^k}{n_N^{k'}}=\frac{2k+1}{2k'+1}\,,
\ee implying that the spectrum in the thermodynamic limit $C(g)=0$ for $g\neq e$.  The ratio of multiplicities approaches that of the regular representation.

\subsubsection*{Free field example} 
Next let us mention briefly some free field theory examples.
We start with the free complex scalar in \(d+1\) dimensions, which has a \(U\left(1\right) \) symmetry. One can compute \(C\left(\alpha\right) \) using path integral methods: starting from \(Z\left(\beta,\alpha\right) = \Tr\left( \text{e}^{-\beta H}\text{e}^{i\alpha Q} \right) \) where \(Q\) is the \(U\left(1\right) \) charge,
\begin{align}
    Z\left(\beta,\alpha\right) = \intop\left[\mathcal{D}\Phi \right]\text{e}^{-\intop_{0}^{\beta}d\tau \intop d^{d}x \left( \left(\partial_{\tau} + i\frac{\alpha}{\beta} \right)\phi^{\dagger}\left(\partial_{\tau} - i\frac{\alpha}{\beta} \right)\phi + \partial_{i}\phi^{\dagger}\partial_{i}\phi + m^{2}\phi^{\dagger}\phi  \right)}.
\end{align}
Using periodicity in time \(\phi\left(\tau+\beta,x\right) = \phi\left(\tau,x\right) \), assuming the system is placed on a \(d\)-dimensional box of volume \(V\),
one can compute \(C\left(\alpha\right) \) and see that it goes to zero unless \(\alpha = 0 \text{ mod }2\pi \),\footnote{As we are in a finite volume the integral is in fact a sum over the momentum modes satisfying some chosen boundary conditions.}
\begin{align}
    C_\mp\left(\alpha\right) = \underset{\beta \rightarrow 0}{\text{lim}} \exp\left[\mp \frac{V}{\beta^{d}} \intop \frac{d^{d}p}{\left(2\pi\right)^{d}} \ln{\left(\frac{ 1+\text{e}^{-2\varepsilon_{\vec{p}}}\mp2\cos{\left(\alpha\right)} \text{e}^{-\varepsilon_{\vec{p}}} }{1+\text{e}^{-2\varepsilon_{\vec{p}}}\mp2\text{e}^{-\varepsilon_{\vec{p}}} } \right)}    \right]\,,
\end{align}
where in fact  $C_-$ is the answer for free bosons and we included also the answer for a free Dirac fermion $C_+$. We defined \(\varepsilon_{\vec{p}} = \sqrt{\vec{p}\cdot \vec{p} + \mu^{2}} \) and \(m = \frac{\mu}{\beta} \) such that $\mu$ and $\vec{p}$ are dimensionless. 
Note the argument inside the \(\ln \geq 1 \) for $C_-$ (and \(\ln \leq 1 \) for $C_+$), whereas it is exactly one for all \(\vec{p}\) only for the identity element, \(\alpha = 0 \). 
The integral itself is finite in the $\beta\to 0$ limit and thus the behavior is determined by the ratio $V/\beta^d$. Thus the argument itself is positive and in the limit of \(\beta\rightarrow 0 \) this quantity is zero unless \(\alpha = 0 \) mod \(2\pi\), where it is exactly one. 
Similarly, one can calculate \(C\left(\alpha\right) \) for free  fermions with the same result.

\subsubsection*{Supersymmetric example}
In gauge theories the computation is more subtle.  Another way to understand the factorization issue discussed above is as follows. Gauging part of the global symmetry, we typically reduce the Hilbert space to gauge-invariant states on one hand and add extra states on the other. The extra states appear concretely {\it e.g.} as configurations of the gauge fields/ twisted sectors. In addition, gauge theories contain natural non-local operators: {\it e.g.} Wilson lines ending on charged local operators. Acting on the vacuum, such non-local operators create new states in the Hilbert space which we need to trace over to compute $C(g)$. Some of the states added due to gauging, {\it e.g.} built from powers of $F_{\mu\nu}$ acting on the vacuum in Maxwell theory, are neutral under global symmetries; others, like the Wilson lines ending on charged operators, might be charged. Instead of trying to analyze these statements in general we will discuss supersymmetric examples.

\

\

Let us consider a supersymmetric version of our statement. 
For concreteness consider ${\cal N}=2$ supersymmetric theory in $4d$.
Instead of regularizing with a Hamiltonian we can consider the following,\footnote{Note that importantly the trace in the Schur index is computed in the radial quantization, {\it i.e.} on ${\mathbb S}^3$.  If one would have considered the usual Witten index \cite{Witten:1982df} $\Tr(-1)^F$ (the spatial manifold being ${\mathbb T}^3$) the index would count imbalance of $(-1)^F$ for vacua with all the higher energies balanced, consistent with our general expectations. The fact that the superconformal index on the other hand receives contributions for arbitrary high dimension operators makes it a reasonable alternative to our general definition of $C(\alpha)$.}
\be\label{eq:defSchurC}
C^S(g) =\lim_{q\to 1}\frac{\Tr (-1)^F \,g\,q^{\Delta-R}}{\Tr (-1)^F \,q^{\Delta-R}}\,.
\ee This is the ratio between the Schur index \cite{Gadde:2011ik,Gadde:2011uv} refined with a global symmetry and one un-refined. We choose this index because it is the simplest one to compute.
Again we would expect this to vanish for non-trivial $g$ and be trivially equal to $1$ for $g=e$.\footnote{Note that the limit of $q\to1$ in the index was studied in the literature. However, usually one also  takes $g\to 1$ with the leading behavior then in $4d$ governed by 't~Hooft anomalies of the theory \cite{Spiridonov:2012ww,Aharony:2013dha,DiPietro:2014bca,ArabiArdehali:2015ybk,Cassani:2021fyv,ArabiArdehali:2023bpq}. An example where a flavor fugacity is kept finite in the limit is the discussion of the Schur index on the {\it second sheet} \cite{Cassani:2021fyv}.}

Let us consider the explicit example of a free hypermultiplet and compute the Schur index,
\be
{\cal I}(q,z)=\Tr (-1)^F \,z^{Q}\,q^{\Delta-R}=\frac{1}{(q^{\frac12}z;q)(q^{\frac12}z^{-1};q)}\,,\qquad (z;q)\equiv \prod_{j=0}^\infty (1-z q^j)\,.\;\;\;\;\;\;
\ee Here $Q$ is the Cartan generator of an $\SU(2)$ global symmetry. The free hypermultiplet has an $\SU(2)$ symmetry fugacity $z$. It is immediately seen that $C^S(z)=0$ unless $z=1$ in which case $C^S(1)=1$. However, we will proceed by showing that the spectrum is regular. 
We can expand the index in irreps of $\SU(2)$,
\be
{\cal I}(q,z)=\sum_{n=0}^\infty C_n(q) \, \chi_n(z)\,,
\ee with $\chi_n(z)$ being the character of the spin $n/2$ irrep.
The index counts (with signs) states in the theory, and thus in principle the coefficients $C_n(q)$ are a measure of how many times the irrep with spin $n/2$ appears. We are viewing the fugacity $q$ as a (UV) regulator such that in the limit $q\to 1$ the
coefficients blow up. However we are interested in ratios of these coefficients, which turn out to be finite. Direct computation gives
\be
C_n(q)= q^{n/2}(1-q^{n+1})\sum_{\ell=0}^\infty\frac{q^{2\ell}}{(q;q)_\ell\,(q;q)_{\ell+n+1}}\,,\qquad (z;q)_\ell\equiv \prod_{j=0}^\ell (1-z q^j)\,,\;\;\;\;\;\;
\ee and this in principle can be expressed in terms of hypergeometric functions. Now we can compute the $q\to 1$ limit to obtain that,
\be\label{eq:Climit}
\lim_{q\to 1}\frac{C_n(q)}{C_m(q)}\,=\frac{n+1}{m+1}\,.
\ee Again we find that the spectrum of irreps (contributing to the Schur index) is regular.
Another way to expand the index is in terms of the Cartan $\U(1)$ of the $\SU(2)$\footnote{
Note that if one  would  count the number of charge $n$ 
states of a free complex scalar in $d$ spatial dimensions with a large hard momentum cut-off $\Lambda$ one would obtain a similar expression with $q\sim \exp\left(- \#/\left( (\Lambda)^dV\right)\right)$ (where $V$ is the volume of space).
},
\be 
{\cal I}(q,z)=\sum_{n=-\infty}^\infty B_n(q) \, z^n\,, \qquad B_n(q)=q^{|n|/2}\sum_{\ell=0}^\infty\frac{q^{\ell}}{(q;q)_\ell\,(q;q)_{\ell+|n|}}\,.
\ee The limit can be readily taken to obtain,
\be\label{eq:Blimit}
\lim_{q\to 1}\frac{B_n(q)}{B_m(q)}=1\,, 
\ee 
as expected as all the representations are one dimensional.\footnote{ A way to prove the statements \eqref{eq:Climit} and \eqref{eq:Blimit} is by noting that the index is 
a theta-function and using modular transformations to show it is given by a $\delta$-function in the limit. The expansion of the $\delta$-function in Fourier modes gives the quoted result.}  

\
\paragraph{\bf Class $\mathcal{S}$ examples:} Next, we generalize this result for interacting theories. Concretely, we consider theories of class ${\cal S}$ with regular punctures \cite{Gaiotto:2009we,Gaiotto:2009hg}. This is a very large class of theories which contains many of the gauge theories with ${\cal N}=2$ supersymmetry as well as conformal theories for which a manifestly ${\cal N}=2$ supersymmetric Lagrangian does not exist \cite{Bhardwaj:2013qia}.\footnote{There can however in some cases be a ${\cal N}=1$ supersymmetric Lagrangian with the extended supersymmetry emerging in the IR \cite{Razamat:2019vfd,Maruyoshi:2016tqk}.}
Theories in class ${\cal S}$ with regular punctures are labeled by an $ADE$ algebra $H$ and a choice of a Riemann surface, {\it i.e} genus $g$ and number of punctures $s$.\footnote{For example $A_1$, $g=0$, and $s=4$ is $\SU(2)$ SQCD with $N_f=4$; $A_1$, $g=1$, $s=1$ is $\SU(2)$ ${\cal N}=4$ SYM with an additional hypermultiplet.} The global symmetry $G$ contains $s$ copies of the $ADE$ group chosen to define the theory.\footnote{We are cavalier here with the global structure and thus do not distinguish the group and the algebra. See \cite{Bhardwaj:2021ojs} for a discussion of the global properties of symmetry groups of class ${\cal S}$. Moreover, the actual symmetry group might enhance. See {\it e.g.} \cite{Gaiotto:2009we,Chacaltana:2010ks,Gaiotto:2012uq}.}
The index then has a uniform behavior,
\be\label{eq:classSindex}
{\cal I}\propto \prod_{j=1}^s\frac1{\prod_{\alpha} (q e^{2\pi i \sum_{i=1}^{{\text{rank}\, H}}\alpha_i\zeta^i_{(j)}};q)}\sum_\lambda C_\lambda(q)\prod_{j=1}^s\chi_\lambda\left(\{e^{2\pi i\zeta^i_{(j)}}\}_{i=1\cdots {\text{rank}\, H}} \right)\,.
\ee Here $\alpha$ are the weights of the algebra $H$; $\lambda$ label finite dimensional irreducible representations of $H$; $\chi_\lambda$ are characters; and $\zeta^j$ are chemical potentials for the Cartan generators of the algebra $H$. The proportionality coefficient does not depend on the chemical potentials for the global symmetry.

Now notice that the index is proportional to factors of the form
\be
\frac1{(q z;q)(q z^{-1};q)}=\sum_{n\in {\mathbb Z}} C_n \,z^n\,,\qquad
C_n =\sum_{\ell=0}^\infty \frac{q^{2\ell+|n|}}{(q;q)_{\ell+|n|}(q;q)_\ell}\,.
\ee In the limit of $q\to 1$ we have as before,
\be
\lim_{q\to 1}\frac{C_m}{C_n} =1\,.
\ee This also implies that if we write
\be
{\cal I}=\sum_{n_1,\cdots, n_{\text{rank}\, H}=-\infty}^\infty C_{\{n^i_1,\cdots, n^i_{\text{rank}\, H}\}_i} \prod_{i=1}^s (z^i_1)^{n^i_1}\cdots (z^i_{{\text{rank}\, H}})^{n^i_{\text{rank}\, H}}\,,
\ee with $z_i=e^{2\pi i \zeta_i}$, then we also expect that
\begin{align}
&\lim_{q\to 1}\frac{C_{\{n^i_1,\cdots, n^i_{\text{rank}\, H}\}_i}}{C_{\{m^i_1,\cdots, m^i_{\text{rank}\, H}\}_i}} =1\,,\\&\left(\forall\,i\,\{n^i_1,\cdots, n^i_{\text{rank}\, H}\}_i\,, \{m^i_1,\cdots, m^i_{\text{rank}\, H}\}_i\,\text{ same N-ality}\right)\,.\;\;\;\; \nonumber
\end{align} The reason is that, if the pre-factor of the sum in \eqref{eq:classSindex} is balanced in the limit, then so is the whole expression, if we look term by term in the sum of \eqref{eq:classSindex}.\footnote{Thus, this argument works rigorously if one truncates the sum over $\lambda$. One should worry about the order-of-limit issues of taking $q\to1$ and performing the sum over $\lambda$. We do not discuss this issue here. However, one can in principle address such questions {\it e.g.} using closed form expressions in terms of Eisenstein series for the flavored Schur index of various  class ${\cal S}$ theories derived in \cite{Pan:2021mrw,Pan:2025vyu} (see also \cite{Beem:2021zvt}) to discuss \eqref{eq:defSchurC}.}
This pre-factor comes from the moment map operators of the theory which sit in the same multiplet as the conserved current: the conserved currents ``regularize'' then the representations. 
We conclude again that  the maximal torus of $G$ has regular representations. 

\subsection{Non-invertible symmetries and regularity}

We expect  the regularity condition to hold also for categorical symmetries. See  \cite{Lin:2022dhv} for a relevant discussion.

We can see this for the simple case of a diagonal RCFT in two dimensions. We  denote the symmetry category by ${\cal C}$.
We consider first the partition function on compact space, {\it i.e.} the torus partition function,
\be
Z=\Tr \, e^{-\beta\, H}= \sum_{h} \chi_h(\tau)\overline{\chi_{\bar h}(\tau)}\,.
\ee Here $\chi_h(\tau)$ are characters of modules labeled by simple objects $h$.  We will take $\tau=i\,\beta$ for simplicity. Now  we use the fact that the Verlinde lines act diagonally,
\be
g\, \chi_h(\tau) =\frac{S_{gh}}{S_{eh}}\,\chi_h(\tau)\,. 
\ee Thus we can write,
\be
Z_g=\Tr \,g\, e^{-\beta\, H}= \sum_{h} 
\frac{S_{gh}}{S_{eh}}\,\chi_h(\tau)\overline{\chi_{\bar h}(\tau)}\,.
\ee We are interested in the  $\beta\to 0$ limit, which is $\tau\to i\, 0^+$. It is useful to use the modular invariance of the partition function and write
\be
\chi(\tau)=S\, \chi(-\frac1\tau)\,,
\ee and then take the limit of $\beta\to 0$,
\be
\chi_h(\tau) = S_{eh}\, C+O(e^{-c/\beta})\,,
\ee where $C$ is a number independent of $h$, and $c$ is a constant. Putting all these together we obtain, in the limit dropping the exponentially suppressed terms,
\be 
C(g)&=&\frac{Z_g}{Z}=\frac{\sum_{h}  \frac{S_{gh}}{S_{eh}}\,\chi_h(\tau)\overline{\chi_{\bar h}(\tau)}}{\sum_{h}  \,\chi_h(\tau)\overline{\chi_{\bar h}(\tau)}}\sim 
\frac{\sum_{h}  \frac{S_{gh}}{S_{eh}} |C|^2 S_{eh} \overline{S_{e\bar h}} }
{\sum_{h}|C|^2 S_{eh} \overline{S_{e\bar h}}}=
\frac{\sum_{h,\bar h} S_{gh} \overline{S_{e\bar h}}}
{\sum_{h,\bar h} S_{eh}\overline{S_{e\bar h}}}=
\frac{\delta_{ge}}{\delta_{ee}}=\delta_{ge}\,.\;\;\;\;\;\;\;\;\;\;
\ee Thus we conclude that for the vast variety of $2d$ diagonal RCFTs the regularity condition,
\be \label{eq:regularitycondition}
   C(g_\alpha) = \lim_{\beta \to 0} \frac{\Tr_\mathcal{H} \left(g_\alpha e^{-\beta H}\right)}{\Tr_\mathcal{H}\left( e^{-\beta H}\right)} =\delta_{g_\alpha,e}\,,
\ee  holds true where $g_\alpha\in {\cal C}$. 
Note that if we define,
\be 
M_{g}(h)\equiv\frac{S_{gh}}{S_{eh}}\,,\qquad M_e(h)=1\,,\qquad
(d_g)^2\equiv |M_g(e)|^2\,.%=\frac{S_{ge}}{S_{ee}}\,,
\ee we have the orthogonality relation,
\be
\frac1{\sum_{g\in {\cal C}} (d_g)^2}\sum_{g\in \cal C} (d_g)^2 M_{h}(g)\overline{M}_{h'}(g)=\delta_{h\,h'}\,,
\ee which generalizes orthogonality of characters for a group.
The quantity $d_g$ is called the {\it quantum dimension} of element $g$. Using these definitions in the limit of $\beta\to 0$ we thus can write that,
\be
C(h) = \frac1{\sum_{g\in {\cal C}} (d_g)^2}\sum_{g\in \cal C} (d_g)^2 \, M_h(g)=\delta_{e\,h}\,.
\ee This is the direct analogue of expression \eqref{eq:Cwithcharacters} for modular fusion categories.

\ 

The statement  \eqref{eq:regularitycondition} in fact holds true for a general fusion category. An argument relies on locality in the following way. If we have a fusion category in a $2d$ QFT then we can discuss symmetry operators, as we did until now, {\it i.e.} topological lines localized in time and extended along space. However, we can also consider topological defects by Wick rotating the lines to be localized in space and extended along the temporal direction. This statement crucially relies on the fact that the operator is locally acting: a consequence of its topological nature.  With this understanding one can compute the torus partition function, $Z_g(\beta)=\Tr_\mathcal{H} \left(g\, e^{-\beta H}\right)$, exchanging the roles of the temporal and the spatial circles using a modular transformation. Then the $\beta\to 0$ limit with the symmetry operator inserted winding the spatial circle is equivalent to $\beta\to \infty$ limit with a topological defect operator winding the temporal circle.

The topological defect can be viewed as a deformation of the Hamiltonian of the system. Let us denote the lowest energy of the deformed Hamiltonian by $\Delta_g$. Then $\Delta_e$ is the lowest energy of the undeformed Hamiltonian. In this modularly transformed setup, in the large $\beta$ limit, the partition function is dominated by the state with lowest energy. In particular we can write
\be
\frac{Z_g(\beta)}{Z_e(\beta)} \sim \frac{c_g\, e^{-\beta\,\Delta_g}}{c_e\,e^{-\beta\,\Delta_e}}\,. 
\ee Thus as we expect $\Delta_{g\neq e}$ to be strictly larger than $\Delta_e$ we conclude that $Z_g(\beta)/Z_e(\beta)\to\delta_{eg}$,
and the statement \eqref{eq:regularitycondition} follows. We stress that for this statement to be true it is crucial to be able to rotate symmetry operators into defect operators, a possibility granted to us by locality. See \cite{Petkova:2000ip,Fuchs:2002cm} for a general and detailed discussion of such partition functions in 2d CFTs.\footnote{ A similar statement was also made in \cite{Harlow:2021trr}. To discuss representations and the analogue of regularity explicitly one would need to go through constructions involving notions of the  tube algebra and the Drinfeld double of a category. As we will not need the details of these constructions for this dicussion we will not delve into it here. See \cite{Lin:2022dhv} for details.}

\subsection{Unitary- but not locally-acting symmetries}\label{sec:unitarybutnotlocal}
We observe, so far, that when symmetries are consistent with locality, the Hilbert space of states is in its regular representation.  In particular this is true for categorical symmetries, as can be seen in the example above and more broadly in \cite{Lin:2022dhv}.  In this section we discuss an example where a unitary symmetry is not acting locally and observe how it violates regularity.  This will highlight the tension between locality and unitarity.

Consider the Ising model which has three simple lines $D$, $\eta$, and the identity $e$ which satisfy the Tambara-Yamagami ${\mathbb Z}_2$ fusion rules,
\be
D^2=e+\eta\,,\qquad \eta^2=e\,,\qquad D\,\eta=\eta\, D=D\,.
\ee Following our general discussion of diagonal RCFTs, we have
\be
C(\eta)=C(D)=0\,,\qquad C(e)=1\,.
\ee The operator $D$ is not invertible but we can construct from it easily invertible operators. We will do this in generality later but here are two examples:
\be
\eta_\pm=\frac1{\sqrt{2}}D\pm\frac{e-\eta}2\,,\qquad \eta_\pm^2=e\,, \qquad \eta\,\eta_+=\eta_+\,\eta\,.
\ee Let us compute the function $C(g)$ for the unitary operators $\eta_\pm$,
\be
C(\eta_\pm)= \frac1{\sqrt{2}} C(D)\pm \frac12\left(C(e)-C(\eta)\right)=\pm \frac12\neq 0. 
\ee Here we can understand the irregularity of representations of the ${\mathbb Z}_2\times {\mathbb Z}_2$ symmetry generated by $\eta$ and $\eta_+$ rather simply. 
Note that from the definition of these operators and the fusion rules it immediately follows that,\footnote{For more on similar restrictions see \cite{Seiberg:2024gek}.}
\be 
(1-\eta)(1-\eta_+)=0\,, \label{eq:etaproj}
\ee which implies that the (odd,odd) sector is missing. 
As the representation of $\eta$ is regular (it is a unitary simple) it follows that the representations of $\eta_+$ cannot be regular. In fact a direct computation shows that $\eta_+$ has even representations thrice as often as odd ones. See Appendix \ref{app:etaplusdefect}. 

To see the non-locality of $\eta_+$, we can can consider its action on local operators.  Let us take the energy operator of Ising $\epsilon$, which is even under $\eta$.  We obtain
\be
\eta_+\,\epsilon \,\eta_+ =-\epsilon\,\eta\,.
\ee It maps it to itself with an $\eta$ line attached.  This can be a consequence of the projector conditions in \eqref{eq:etaproj}.  We can compute the eigen-operators of the system associated to the $\epsilon$ as
\be
{\cal O}^{+\pm}=\frac{\epsilon\pm\eta_+\,\epsilon\,\eta_+}2= \epsilon\,\frac{1\mp\eta}2\,.
\ee The operator involves as usual a projector and thus is not local.  
The ${\mathbb Z}_2\times {\mathbb Z}_2$ group structure leads to selection rules for operators with well-defined charges. As these involve linear combinations of non-local operators we get relations between correlators of topological and local operators. The fact that the eigen-operators are non-local has interesting implications for dynamical systems which we will explore in the future.  Recently, we have seen these constraints in various analyses of non-invertible and categorical symmetries \cite{Komargodski:2020mxz,Bartsch:2023pzl,Bartsch:2023wvv,Bhardwaj:2023wzd,Bhardwaj:2023ayw,Cordova:2024iti}. %\ib{cite Zohar's, Clay's, and other similar papers here.} 

\ 

Next we will develop a systematic understanding of the relation between irregularity and non-invertible symmetries in the unitary approach.

\section{A unitary perspective and irregularity of representations}
\label{sec:unitaryapproach}

We have observed so far that locally acting symmetries have particular regularity in their action on the Hilbert space of states.  As importantly, this property is also satisfied by categorical symmetries, since the operators they encode are topological.  On the other hand, categorical symmetries are generically non-invertible, which naively seems in tension with Wigner's theorem in quantum mechanics where symmetry operators are unitary.  In this section we will consider unitarity as paramount and explore properties of the trace observable $C(g)$.  Without the imposition of locality this object need not vanish.  {\it The value of $C(g)$ is a measure of non-locality of the action of a unitary symmetry operator $U(g)$ in quantum field theory.} In this section we characterize some of its important properties.  

To study the unitarity perspective, we will consider a system with a prescribed symmetry, which can be categorical, and explore the space of unitary operators it implies.  
This will be the set of unitary operators that commute with the Hamiltonian of the system. It necessarily builds the structure of a group $G$. Importantly, we assume that $G$ is not spontaneously broken: the  vacuum is a singlet of $G$. We do not assume here that the operators associated to $g\in G$ correspond to topological defects where they would enjoy locality properties.  In particular the action of the operator can be non-local.  

For every group element $g\in G$, we define
\be
\label{eq:regularityconditionG}
    B(g)=\left|C(g)\right|^2\,.
\ee We consider unitary $g$'s that can be continuously deformed to the identity operator.
However, now $C(g)$ and thus $B(g)$ can be non-vanishing even for $g\neq e$. 

\

Let us give an example. Consider a quantum system with a categorical symmetry, {\it i.e.} a symmetry from the local perspective, which is just the group ${\mathbb Z}_2$  generated by  $\eta$. Then the most general unitary operator will be given by
\be
g(\theta)=\cos\,\theta\, e+i\, \sin\,\theta \, \eta \,.
\ee 

Now, using the fact that for local symmetries $C(\eta)=0$, we immediately derive that
\be
C(g(\theta))=\cos\,\theta\,,\qquad B(\theta)=\cos^2\theta\,.
\ee The reason these quantities are not zero in general is that they involve a linear combination of the unit operator with other operators. In particular we notice that $B((2n+1)\pi/2)=0$ as $g(\pi/2)=(-1)^n i\, \eta$ and $B(n\,\pi )=1$ as $g(0)=(-1)^n\,e$. For other values of $\theta$ $B(\theta)$ is neither one nor zero.

In general if we have some finite categorical symmetry generated by $g_\ell$, one can immediately build a set of unitary operators $\exp\left(i\alpha_+ (g_\ell+g^{\dagger}_\ell)\right)$ and $\exp\left(\alpha_-(g_\ell-g^{\dagger}_\ell)\right)$. Thus,  the set $G$ will be in general a Lie group. If an underlying categorical structure is generated by a finite number of simple objects it will be a finite-dimensional Lie group. For simplicity, we will assume that this is the case from now on.

\ 

As another example consider a theory with the Fibonacci category as its symmetry from the local perspective. The Fibonacci category is generated by a hermitian operator $W$ such that,
\be
W^2=e+W\,.
\ee A general unitary operator is then given (up to an over-all phase) by
\be 
g(\theta)=e^{i \theta W}=\frac{3+\sqrt{5}+2e^{i\sqrt{5}\theta}}{5+\sqrt{5}}\,e+\frac{e^{i\sqrt{5}\theta}-1}{\sqrt{5}}W
%\frac1{\sqrt{1+4\cos^2\theta}}-\frac{2\cos\,\theta}{\sqrt{1+4\cos^2\theta}}e^{i\,\theta}\,W\,.
\ee Thus
\be \label{eq:fobonacciB}
B(\theta)=\frac{3+2\cos\sqrt{5}\theta}{5}\,.
\ee Note that $B(\theta)$ does not vanish, it is equal to one for $\theta=0$ and thus $g=e$ and has a minimum at $\theta=\frac{\pi}{\sqrt{5}}$ with
\be
g\left(\frac{\pi}{\sqrt{5}}\right)=\frac{1-2W}{\sqrt{5}}\,,\qquad g\left(\frac{\pi}{\sqrt{5}}\right)^2=e\,,\qquad B\left(\frac{\pi}{\sqrt{5}}\right)=\frac15\, .
\ee This generates a ${\mathbb Z}_2$ symmetry. In fact this symmetry is a rather natural one, as we will  discuss soon.

\

\subsection{The function $B(g)$ as a measure of irregularity}\label{sec:propertiesofB}

\noindent We claim that $B(g)$  satisfies the following properties:
\begin{enumerate}
    \item The function $B$ is bounded,\be 0\leq B\leq 1\,.\ee
    \item If the categorical symmetry from the local perspective has $n$ simple defects, the group manifold of $G$ is $n-1$ dimensional. 
    \item $B$ may have minima with $B=0$. Some of these correspond to locally acting unitary operators. These unitary operators along with the identity operator form a group which can be identified with topological operators. This is conversely true, all invertible topological symmetry operators (except the identity) will satisfy $B=0$.
    \item The maxima with $B=1$ correspond to the unit operator (or a phase times a unit operator).
    \item Some of the critical points of $B$ with $0<B<1$ correspond to special unitary operators.  The origin of these critical points can be understood from gauging operators of the symmetry category.  
    More specifically, consider a subcategory ${\cal C}'$ of the symmetry category with simple objects $g_a$ with quantum dimensions $d(g_a)$. We can build a projector,
    \be
    P_{{\cal C}'}= \frac{\sum_{g_a\in{\cal C}'} d(g_a)\,g_a}{\sum_{g_a\in{\cal C}'} d(g_a)^2}\,,
    \ee and a ${\mathbb Z}_2$ generator,
    \be
    \eta_{{\cal C}'} =e-2P_{{\cal C}'}\,.
    \ee These unitary elements lead to critical points with $0<B<1$ for various possible operators $\eta_{{\cal C}'}$ with
    \be\label{eq:valueofBatcriticalpoints}
    B=\left(1-\frac2{\sum_{g'_a\in{\cal C}'} d(g'_a)^2}\right)^2\,.
    \ee
\end{enumerate}
The proofs of these statements are summarized in Section \ref{sec:simpleproofs}. These statements are true if the symmetries act on the Hilbert space non-projectively. If projective phases are present then some of these statements will be modified manifesting the {\it anomaly}: we will see an example later on.

First let us see how these statements play out in our two simple examples. In the ${\cal C}={\mathbb Z}_2$ case we have a one dimensional $\U(1)$ group manifold as expected. The maximum corresponds to $e$ and the minimum with $B=0$ corresponds to $\eta$.
For ${\mathbb Z}_2$ the gauge operator is identical to $\eta$:
\be
P_{{\cal C}}=\frac{e+\eta}{2}\;\;\to \;\; \eta_{{\cal C}}=e-2\,\frac{e+\eta}{2}=\eta\,.
\ee Note that we can also think of the maximum as the gauging projector for the trivial group ($P_{{\cal C}'}=e$ and $\eta_{{\cal C}'}=e-2e=-e$).

For the Fibonacci ${\cal C}$ we again have a one-dimensional $\U(1)$ group manifold as expected. The maximum corresponds to $e$ and the minimum is {\it not} with $B=0$. This reflects the fact that the fusion category does not have a simple defect that is also unitary. However, the existence of this minimum seems still to encode the information about the simple line $W$.

Let us assume we are given the function $B(\theta)$ of \eqref{eq:fobonacciB} without the knowledge of its origin from the fusion category.  Could we derive the fusion rules? Since we have a one-dimensional group manifold, we deduce that we have two simple elements, one of which has to be $e$ and the second one we denote by $W$. The minimum we interpret as coming from the gauging projector and thus
\be
B(\frac{\pi}{\sqrt{5}})=\frac15=\left(1-\frac{2}{1+d(W)^2}\right)^2\;\;\to\;\; d(W)=\frac{1+\sqrt{5}}2\;\text{or}\;\frac{-1+\sqrt{5}}2\,.
\ee Since the quantum dimensions should be consistent with the fusion the above values have to satisfy,\footnote{Note that $N^e_{WW}=1$ because of \eqref{eq:Nidentities}.}
\be
d(W)^2=1+N^W_{WW} d(W)\,,
\ee and thus we deduce that if $d(W)=\frac{1+\sqrt{5}}2$ then $N^W_{WW}=1$ and if $d(W)=\frac{-1+\sqrt{5}}2$ then $N^W_{WW}=-1$. However, since the fusion coefficients are non-negative integers only the former is a valid option. We thus recover the  Fibonacci fusion rules by studying properties of $B(W)$.

\ 

Note that in both examples the group manifold is $\U(1)$. However, the local structure is different and that fact is encoded in the properties of $B(\theta)$.

\ 

Let us next consider the consequences of the statements above more generally. First, given any quantum theory we 
can find the set of all unitary symmetries and compute the function $B$. We can then try to derive the categorical structure from the properties of $B$. First, the dimension of ${\cal C}$ is given by the dimension of the group manifold plus one. Then we can study the set of  zeroes of $B$. Some, but not all, zeroes correspond to unitary simples. We can look at the set of unitary operators with $B=0$ which form the structure of a group. The ones who are part of ${\cal C}$ moreover lead to gauging projectors and thus to critical points with values of $B$ as in \eqref{eq:valueofBatcriticalpoints}. Once the invertible part of ${\cal C}$ is identified, one can proceed to study the critical points and derive constraints on ${\cal C}$.
We expect that it will be possible to determine quantum dimensions from the critical points and to use these to determine (or at least constrain) the fusion coefficients.
Note that since we will have gauging projectors for any sub-category there is a lot of information encoded in $B$.\footnote{An example of a general expectation is as follows. If there are unitary categorical symmetries generated by $\hat g_\alpha$ there have to be gauging projector critical points for every subgroup. For example, one could have a ${\mathbb Z}_{k_\alpha}$ subgroup with $k_\alpha$ the order of $\hat g_\alpha$. Then there should be critical points with $B=\left(\frac{k_\alpha-2}{k_\alpha}\right)^2$.
} 

There are however  caveats that we want to stress. First, we have assumed that the local symmetry is categorical and that the unitary symmetry can be expressed in terms of linear combinations of simples. This is a non-trivial assumption. If there are unitary operators which are not expressible in terms of the simples our general expectations do not hold. In such a case one should look for subgroups of $G$ for which the constraints from the general expectations above can be consistently resolved. 

\ 

\noindent Let us next derive the statements $1-5$ about the function $B(g)$.

\

\subsection{Derivations of properties of $B(g)$}\label{sec:simpleproofs}

We  first note that given a fusion category the quantum dimensions satisfy the following properties \cite{EtingofGelakiNikshychOstrik2015}
%\ib{Need a citation here},
\be\label{eq:Nidentities}
d(g_a)d(g_b)=\sum_c N^c_{ab}\, d(g_c)\,,\qquad d(g_{\bar c})=d(g_c)\,,\qquad 
N^c_{ab}=N^{\bar b}_{{\bar c}a}=N^{\bar a}_{b{\bar c}}\,.
\ee

\ 

First let us assume that we have a category ${\cal C}$ with $n$ simples. The general unitary operator can be written as a linear combination of the simples,
\be\label{eq:gwithsimples}
g = \sum_\alpha {m}^\alpha\, g_\alpha\,,
\ee with coefficients $m^\alpha$ satisfying
\be
e=\left(m^\alpha g_\alpha\right)\left((m^{\bar\beta})^\dagger g_{\bar \beta}\right) =m^\alpha {N^\gamma}_{\alpha\bar\beta}(m^{\bar\beta})^\dagger\, g_\gamma\,.
\ee Thus
\be 
m^\alpha {N^e}_{\alpha\bar\beta}(m^{\bar\beta})^\dagger=1\,,\qquad 
m^\alpha {N^{\gamma\neq e}}_{\alpha\bar\beta}(m^{\bar\beta})^\dagger=0\,. \label{eq:Cconst}
\ee These are $n$ equations  for $n$ complex numbers $m^\alpha$. As we can always multiply a solution for $m^\alpha$ by an $\alpha$-independent phase the dimension of the space of solutions  thus is  expected to be $n-1$.

\

Let us next show that
\be
    P_{{\cal C}'}= \frac{\sum_{g'_a\in{\cal C}'} d(g_a)\,g_a}{\sum_{g'_a\in{\cal C}'} d(g_a)^2}\,
    \ee
is a projector. Taking into account \eqref{eq:Nidentities} we compute,
\be
&&\left(\sum_a d(g_a)g_a\right)^2=\sum_{a,b} d(g_a)d(g_b)g_a g_b=
\sum_{a,b,c} d(g_a)d(g_b)N^{c}_{ab} g_c=\\
&&\sum_{\bar a,b,\bar c} d(g_a)d(g_b)N^{\bar c}_{{\bar a}b} g_{\bar c}=
\sum_{a,b,c} d(g_a)d(g_b)N^{a}_{bc} g_{\bar c}=\sum_{b,c} d(g_b)d(g_b)d(g_c) g_{\bar c}=\nonumber\\
&&\left(\sum_b d(g_b)^2\right)\,\left(\sum_c d(g_c) g_c\right)\,,\nonumber
\ee which completes the proof. 

\ 

Let us show that $\eta_+\equiv e-2P_{{\cal C}'}$ are critical points of $B(g)$.  A general unitary element in terms of the simples is written in \eqref{eq:gwithsimples}.
Using this we also can write $B=m^e(m^e)^\dagger$. For the $\eta_+$
we have\footnote{Here we write only $m^\alpha$ for elements of ${\cal C}'$ as those in ${\cal C}\setminus{\cal C}'$ are zero. }
\be
m^\alpha=\delta^\alpha_e-\frac{2d(g_\alpha)}{\sum_{g_\rho\in{\cal C}'} d(g_\rho)^2}\,.
\ee We note that these $m^\alpha$ are real. Next we perturb, $m^\alpha\to m^\alpha+\delta m^\alpha$ and expand \eqref{eq:Cconst} to linear order,
\be 
\left(\delta^\alpha_e-\frac{2d(g_\alpha)}{\sum_{g_\rho\in{\cal C}'} d(g_\rho)^2}\right){N^{\gamma}}_{\alpha\bar\beta}(\delta m^{\bar\beta})^\dagger+
\delta m^\alpha {N^{\gamma}}_{\alpha\bar\beta} 
\left(\delta^{\bar \beta}_e-\frac{2d(g_{\bar\beta})}{\sum_{g_\rho\in{\cal C}'} d(g_{\rho})^2}\right)=0\,.
\ee Using the fact that ${N^{\gamma}}_{\alpha e} = {N^{\gamma}}_{e\alpha}=\delta^\gamma_\alpha$ this can be written as
\be
\left(\delta^\gamma_\beta-\frac{2d(g_{\beta})d(g_{\gamma})}{\sum_{g_\rho\in{\cal C}'} d(g_{\rho})^2}\right)\, \text{Re }( \delta m^\beta) =0\,.
\ee From here we immediately deduce by multiplying with $d(g_\gamma)$ and summing over $\gamma$ that
\be
\left(d(g_{\beta})\text{Re }( \delta m^\beta)\right)\left(1-2\right)=0\;\; \to \;\; d(g_{\beta})\text{Re }( \delta m^\beta)=0\,,
\ee
and thus $\text{Re }( \delta m^\beta) =0$. From here in  particular it follows that
\be\label{eq:Bsaddle}
\delta B= 2C^e \text{Re }( \delta m^e)=0\,,
\ee and thus the gauging  ${\mathbb Z}_2$ projector generated by $\eta_+$ is a critical point of $B$. From here we also can deduce that the coordinates near these critical points are given by turning on small imaginary values for $m^\alpha$. 

Note that the same proof holds if instead of $\eta_+=e-2P_{{\cal C}'}$ we consider
$\eta_-\equiv\eta-2P_{{\cal C}'}$ such that $\eta^2=e$. Thus, whenever the categorical symmetry has ${\mathbb Z}_2$ subgroups we should see additional critical points. For example, in the next section we will observe this for the ${\mathbb Z}_2$ Tambara-Yamagami, ${\mathbb Z}_2\times {\mathbb Z}_2$, and the $Rep(S_3)$ categories. Here we have
\be
m^\alpha=\delta^\alpha_\eta-\frac{2d(g_\alpha)}{\sum_{g_\rho\in{\cal C}'} d(g_\rho)^2}\,, 
\ee and perturbation gives to linear order,
\be \label{eq:Z2projector}
\left(\delta^\alpha_\eta-\frac{2d(g_\alpha)}{\sum_{g_\rho\in{\cal C}'} d(g_\rho)^2}\right){N^{\gamma}}_{\alpha\bar\beta}(\delta m^{\bar\beta})^\dagger+
\delta m^\alpha {N^{\gamma}}_{\alpha\bar\beta} 
\left(\delta^{\bar \beta}_\eta-\frac{2d(g_{\bar\beta})}{\sum_{g_\rho\in{\cal C}'} d(g_{\rho})^2}\right)=0\,,
\ee multiplying which by $d(g_\gamma)$ and summing over $\gamma$ gives, denoting ${\cal D}^2=\sum_{g_\rho\in{\cal C}'} d(g_\rho)^2$,
\be
\left(d(\eta)d(\bar\beta)-\frac{2d(g_{\bar\beta}){\cal D}^2}{{\cal D}^2}\right)(\delta m^{\bar\beta})^\dagger+
\delta m^\alpha 
\left(d(\eta)d(\alpha)-\frac{2d(g_{\alpha}){\cal D}^2}{{\cal D}^2}\right)=0\,,
\ee leading to $d(g_{\beta})\text{Re }( \delta m^\beta)$ again as $d(\eta)=1$. Then \eqref{eq:Z2projector} becomes
\be
{N^{\gamma}}_{\eta\bar\beta}(\delta m^{\bar\beta})^\dagger+
\delta m^\alpha {N^{\gamma}}_{\alpha\eta}=0\,.
\ee Finally, take $\gamma=\eta$ and use the fact that ${N^{\eta}}_{\eta\bar\beta}=\delta^e_{\bar\beta}$ because $\eta$ is a ${\mathbb Z}_2$ generator, to deduce that $\text{Re }( \delta m^e)=0$ leading to \eqref{eq:Bsaddle} and the conclusion that  $\eta_-=\eta-2P_{{\cal C}'}$ is a critical point of $B$. Moreover, we can show that $\eta\eta_+=\eta_-$,
\be
&&\eta\eta_+=\eta-\frac{2d(g_\alpha)\eta g_\alpha}{\sum_{g_\rho\in{\cal C}'} d(g_\rho)^2}=
\eta-\frac{2d(g_\alpha){N^\gamma}_{\eta\alpha} g_\gamma}{\sum_{g_\rho\in{\cal C}'} d(g_\rho)^2}=
\eta-\frac{2d(g_\alpha){N^{\bar\alpha}}_{\bar\gamma\eta} g_\gamma}{\sum_{g_\rho\in{\cal C}'} d(g_\rho)^2}=
\\
&&\;\;\;\eta-\frac{2d(g_\gamma)d(\eta)g_\gamma}{\sum_{g_\alpha\in{\cal C}'} d(g_\alpha)^2}=\eta_-\,,\nonumber
\ee where we used $d(\eta)=1$ and $d(g_\rho)=d(g_{\bar\rho})$.

\

In the next section we will discuss briefly several examples of the interplay between the categorical and the unitary approach, with each example shedding light on some of the features of the general picture.

\

\section{Examples}\label{sec:examples}

Let us discuss several examples of symmetry categories and explicit computations of $B$.

\subsection{Tambara-Yamagami ${\mathbb Z}_2$}

We already briefly mentioned this example but now let us discuss it in detail.
A canonical example of a categorical symmetry ${\cal C}$ is that of Tambara-Yamagami ${\mathbb Z}_2$ category,
\be\label{eq:TYfusion}
D^2=e+\eta\,,\qquad \eta^2=e\,,\qquad D\,\eta=\eta\, D=D\,.
\ee realized as the symmetry category of the Ising $2d$ CFT. Note that $\eta$ is a unitary ${\mathbb Z}_2$ symmetry while $D$ is non-unitary (not invertible) but is hermitian. A general unitary operator here can be written as,
\be \label{eq:TYZ2group}
&&g(\theta,\phi)=e^{i\theta\,\eta}\,e^{i\phi\, D}=\\
&&\;\;\;\;\frac12\left(\cos\sqrt{2}\phi \,e^{i\theta} +e^{-i\theta}\right)\,e+
\frac12\left(\cos\sqrt{2}\phi \,e^{i\theta}-e^{-i\theta}\right)\,\eta+\frac{i}{\sqrt{2}}\sin\sqrt{2}\phi e^{i\theta}\, D\,.\nonumber
\ee  The group manifold is $\U(1)^2$ and we have,
\be
B(\theta,\phi)=\frac{1+\cos^2\sqrt{2}\phi+2\cos2\theta \cos\sqrt{2}\phi}4\,.
\ee Let us analyze the critical points of $B$. It has minima at,
\be
(\theta,\phi)=(0,\pm\frac{\pi}{\sqrt{2}})\,,\;(\frac{\pi}{2},\pm\sqrt{2}\pi)\,.
\ee All of the minima have $B=0$ and the value of the group element $g$ is $\pm \eta$ and thus correspond to the generator of the ${\mathbb Z}_2$ subgroup of ${\cal C}$.  The maxima are located at,
\be
(\theta,\phi)=(\frac{\pi}{2},\pm\frac{\pi}{\sqrt{2}})\,,\;(0,\pm\sqrt{2}\pi)\,,
\ee and all have $B=1$ with $g= e$ or $g=-i \, e$ as expected.
Finally the saddle points are at,
\be
(\theta,\phi)=(\pm \frac{\pi}4,\pm\frac{\pi}{2\sqrt{2}})\,.
\ee (The signs here are not correlated.) The value of $B$ at all saddle points is equal to $\frac14$. Again let us perform the exercise of deriving the fusion coefficients from this information. We know that we have three simples because the dimension of the group manifold is two. One is $e$ and another is a group that we will call $\eta$. We know that this is a ${\mathbb Z}_2$ as if it as a bigger group we should have found more group elements with $B=0$. As such both $e$ and $\eta$ have quantum dimension $1$. Thus as $B=\frac14$ for the saddle we write
\be
\frac14=\left(1-\frac{2}{1+1+d(D)^2}\right)^2\,,
\ee where we denoted the extra element of ${\cal C}$ by $D$. 
We thus obtain that $d(D)=\sqrt{2}$: the equation has more solutions but these are either imaginary or negative, which a quantum dimension cannot satisfy. Next we consider the fusion
\be
&&d(D)^2=1+N^D_{DD} d(D)+N^\eta_{DD}\,,\qquad
d(D)d(\eta)=N^D_{D\eta}d(D)+N^\eta_{D\eta}\,,\qquad\\
&&d(\eta)d(D)=N^D_{\eta D}d(D)+N^\eta_{\eta D}\,.\nonumber
\ee We know from the group law of $\eta$ that $N^D_{\eta\eta}=0$ and thus also $N^\eta_{\eta D}=N^\eta_{D\eta}=0$. Plugging $d(\eta)=1$, $d(D)=\sqrt{2}$ and  since $D$ is hermitian\footnote{Otherwise we would have obtained two different simples in contradiction to our findings.}  this is solved to give
\be
N^D_{\eta D}=N^D_{D\eta}=N^\eta_{DD}=1\,, \qquad N^D_{DD}=0\,. 
\ee This is precisely the fusion of the ${\mathbb Z}_2$ Tambara-Yamagami category. 

Finally let us mention that the group elements at the critical  points with $0<B< 1$ are
\be
\frac{e-\eta}2\pm \frac1{\sqrt{2}}D\,,\qquad e^{\frac{i\pi}4}\left(\frac{e-\eta}2\pm \frac1{\sqrt{2}}D\right)\,.
\ee Note first that the overall phase is not essential. Moreover, we can write
\be
e-2\frac{e+\eta+\sqrt{2}D}{1+1+2}=\frac{e-\eta}2- \frac1{\sqrt{2}}D\,,
\ee which matches one class of the solutions. 
The second class is obtained by taking $D\to -D$: note that this does not change the fusion rules at all and thus all the statements follow in the same way. We learn thus that if there is some transformation of the simples by phases that does not change the fusion coefficients we can also build a gauging ${\mathbb Z}_2$ for these. The group elements at these critical points correspond (up to phases) to the $\eta_{\pm}$ ${\mathbb Z}_2$ symmetries defined and discussed in Section \ref{sec:unitarybutnotlocal}.\footnote{
We can generalize much of the the discussion in this section rather straightforwardly to  ${\mathbb Z}_k$ Tambara-Yamagami,
$D^2=\sum_{\ell=1}^k g^\ell$.
 Here $g$ is generator of the ${\mathbb Z}_k$ symmetry. For example, defining a ${\mathbb Z}_2$ generator
$\eta_+=D+e-\frac{1}k\sum_{\ell=1}^k g^\ell$,
 we would deduce that $B({\eta_+})=\left(\frac{k-1}{k}\right)^2$\,.
}

\ 

\subsection{${\mathbb Z}_3$ and ${\mathbb Z}_k$}
Another example corresponds to the \(\mathbb{Z}_{3} \) group containing three elements, \(e,\eta,\eta^{2}\):
\begin{align}
    \eta^{3}=e,\qquad \eta^{2}=\eta^{-1}.
\end{align}
Note that \(\eta \) and \(\eta^{2}\) are unitary, \(\eta^{\dagger} = \eta^2=\eta^{-1} \). A general unitary operator built from them is
\begin{align}
   \begin{aligned}
        g\left(\theta,\phi\right) &= \text{e}^{i\theta}\text{e}^{i\theta\left(\eta+\eta^{2}\right)} \text{e}^{\phi \left(\eta-\eta^{2}\right)}\\
        &= \frac{1}{3}\left(\text{e}^{i 3\theta} + 2\cos{\left(\sqrt{3}\phi \right)} \right)\\
        &+ \frac{1}{3}\left(\text{e}^{i 3\theta} - \cos{\left(\sqrt{3}\phi \right)} + \sqrt{3}\sin{\left(\sqrt{3}\phi \right)} \right)\eta + \frac{1}{3}\left(\text{e}^{i 3\theta} - \cos{\left(\sqrt{3}\phi \right)} - \sqrt{3}\sin{\left(\sqrt{3}\phi \right)} \right)\eta^{2},
   \end{aligned}
\end{align}
where we include an overall phase in the definition \(\text{e}^{i\theta} \) for convenience. The \(B\left(\theta,\phi\right) \) function is given by
\begin{align}
    B\left(\theta,\phi\right) = \frac{1+4\cos^{2}\left(\sqrt{3}\phi \right) + 4 \cos{\left(\sqrt{3}\phi\right)\cos{\left(3\theta\right)}}}{9}.
\end{align}
The group manifold is $\U(1)^2$ and \(g\left(\theta,\phi\right) = g\left(\theta+\frac{2\pi}{3}n,\phi+\frac{2\pi}{\sqrt{3}}m\right) \) for  integer \(n,m\):  we can explore the critical points  of \(B\) within the range \(0\leq \theta<\frac{2\pi}{3} \), \(0\leq \phi<\frac{2\pi}{\sqrt{3}} \). The function $B$ has minima at
\begin{align}
    \left(\theta,\phi\right) = \left(0,\frac{2\pi}{3\sqrt{3}}\right), \left(0,\frac{4\pi}{3\sqrt{3}}\right), \left(\frac{\pi}{3},\frac{\pi}{3\sqrt{3}}\right),\left(\frac{\pi}{3},\frac{5\pi}{3\sqrt{3}}\right),
\end{align}
which all correspond to group elements \(\pm \eta \) and \(\pm \eta ^{2} \) with \(B = 0\). The maxima are located at
\begin{align}
    \left(\theta,\phi\right) = \left(0,0\right),\left(\frac{\pi}{3},\frac{\pi}{\sqrt{3}}\right),
\end{align}
which correspond to the identity element \(\pm 1\) with \(B=1\). The saddle points are at
\begin{align}
    \left(\theta,\phi \right) = \left(0,\frac{\pi}{\sqrt{3}}\right), \left(\frac{\pi}{6},\frac{\pi}{2\sqrt{3}} \right),  \left(\frac{\pi}{6},\frac{\sqrt{3}\pi}{2} \right), \left(\frac{\pi}{3},0\right),\left( \frac{\pi}{2},\frac{\pi}{2\sqrt{3}} \right), \left( \frac{\pi}{2},\frac{\sqrt{3}\pi}{2} \right),
\end{align}
where \(B = \frac{1}{9}\) for all of these saddle points. 
The group elements associated with these points 
can be derived from \(\pm i \frac{1-2\eta-2\eta^{2}}{3} \), such that the others correspond to redefining the group element \(\eta \rightarrow \text{e}^{\pm i\frac{2\pi}{3}}\eta \), such that \(\eta^{3}=1 \):
\begin{align}
    \pm i \frac{1-2\eta-2\eta^{2}}{3},\qquad \pm i \frac{1-2\text{e}^{i\frac{2\pi}{3}} \eta-2 \text{e}^{-i\frac{2\pi}{3}}\eta^{2}}{3},\qquad  \pm i \frac{1-2\text{e}^{-i\frac{2\pi}{3}} \eta-2 \text{e}^{i\frac{2\pi}{3}}\eta^{2}}{3}.
\end{align}

\

Let us next comment on  the general ${\mathbb Z}_k$ case. We denote by $g$ the generator of the group. Then
\be\label{eq:Zkg}
&k=2m+1:&\;\; g(\theta_i,\,\phi_i)=\prod_{j=1}^{m}e^{i\theta_j(g^j+g^{-j})}e^{\phi_j(g^j-g^{-j})}=\prod_{i=1}^{2m} e^{i\chi_j g^j}\,,\\
&k=2m:&\;\;g(\theta,\,\theta_i,\,\phi_i)=e^{i\theta g^m}\prod_{j=1}^{m-1}e^{i\theta_j(g^j+g^{-j})}e^{\phi_j(g^j-g^{-j})} =\prod_{i=1}^{2m-1} e^{i\chi_j g^j}\,,\nonumber
\ee such that $\chi_j=\theta_j-i\,\phi_j$ and $\chi_j=\overline{\chi_{k-j}}$. It is easy to identify the gauging ${\mathbb Z}_2$. Let us consider an order $n$ subgroup 
generated by $g^{k/n}$, then the gauging ${\mathbb Z}_2$ for this is given by taking all $\chi$ for power of $g^{k/n}$ to be equal to $\frac{\pi}n$ and zero otherwise. Thus the relevant element is (multiplying by $e^{\frac{\pi i\,e}k}$ for convenience) 
\be
g(\chi)=e^{\frac{\pi i}n\sum_{\ell=0}^{n-1}g^{\frac{k\,\ell}{n}}}=e-\frac2n\sum_{\ell=0}^{n-1}g^{\frac{k\,\ell}{n}}\,.
\ee Let us now consider a small perturbation,
\be
\chi_{\frac{k\,\ell}{n}}=\frac{\pi}n+\delta_{\frac{k\,\ell}{n}}\,,\qquad \chi_{s\neq \frac{k\,\ell}n}=\delta_s\,,
\ee and then to leading order in the perturbation,
\be
\prod_{i=1}^{k-1} e^{i\chi_j g^j}\;\to\; 
\left(e-\frac2n\sum_{\ell=0}^{n-1}g^{\frac{k\,\ell}{n}}\right)\left(e+i\, \sum_{\ell=1}^{k-1}\delta_{\ell}g^\ell\right)\,.
\ee The coefficient of the identity is
\be
C^e=\frac{n-2}n-\frac{2i}n \sum_{\ell=1}^{n-1}\delta_{\frac{k\,\ell}{n}}\;\;\to\;\;
B=|C^e|^2=\frac{(n-2)^2}{n^2}+O(\delta^2)\,,
\ee which does not have linear terms in the deformation, and thus we deduce that the gauging ${\mathbb Z}_2$ for any subgroup are critical points. We used above the unitarity condition $\chi_j=\overline{\chi_{k-j}}$ and thus $\sum_{\ell=0}^{n-1}\delta_{\frac{k\,\ell}{n}}$ is real.

\ 

\subsection{${\mathbb Z}_2\times {\mathbb Z}_2$ with and without a 't~Hooft anomaly} 

Let us consider a categorical symmetry which is just the ${\mathbb Z}_2\times {\mathbb Z}_2$ group but realized projectively on the Hilbert space,
\be
g^2=h^2=e\,,\qquad g\,h=-h\,g\,.
\ee Note that because of the anomaly this case does not fit precisely our general discussion: the appearance of the minus sign in the fusion rule goes beyond a categorical definition of fusion. We could in principle cure this by centrally extending with an element $c$, which would double the number of simples. The group we would obtain is the dihedral one with eight elements. However, this is not natural from the point of view of classifying all the possible unitary operators commuting with the Hamiltonian. We thus will proceed as usual and will observe a new feature appearing.

With the group laws above we can define,
\be
t_1\equiv\frac{g}2\,,\qquad t_2\equiv\frac{h}2\,,\qquad 
t_3\equiv\frac{i\,hg}2\,,
\ee such that,
\be
[t_i,\, t_j]=i\epsilon_{ijk} t_k\,.
\ee This is just the usual Clifford algebra construction of a spin $\frac12$ representation of $\SU(2)$. 
The group manifold is the $\SU(2)$ and the general unitary operator can be then written as
\be 
g(\theta^1,\theta^2,\theta^3)=e^{i\theta^k t_k}= \cos\frac{\sqrt{\theta^i\theta_i}}{2}\,e+\frac{2i\sin\frac{\sqrt{\theta^i\theta_i}}{2}}{\sqrt{\theta_i\theta^i}}\,\theta^k\,t_k\,.\nonumber
\ee From here,
\be
B(\theta^1,\theta^2,\theta^3)=\cos^2\frac{\sqrt{(\theta^1)^2+(\theta^2)^2+(\theta^3)^2}}{2}\,.
\ee The critical points satisfy
\be
\theta^k\,\frac{\sin\sqrt{(\theta^1)^2+(\theta^2)^2+(\theta^3)^2}}{\sqrt{(\theta^1)^2+(\theta^2)^2+(\theta^3)^2}}=0\,.
\ee We have a maximum as usual at $\theta_k=0$ corresponding to $e$. We have minima for
\be\label{eq:sphereconditions}
(\theta^1)^2+(\theta^2)^2+(\theta^3)^2=n^2\pi^2\,,
\ee for some $n\in {\mathbb Z}$. We do not have any saddle points.

Let us as usual assume this structure of $B$ and reverse engineer the category. We expect four simples as the dimension of the group manifold is three. One of these is the identity $e$. Since we do not have any saddle points or other critical points with $0<B<1$ the three remaining simples must each square to $e$. Let us denote these as $h_i$: $h_i^2=e$. Since all the simples are invertible we have a group. The only group with these properties is  ${\mathbb Z}_2\times {\mathbb Z}_2$ either with or without an anomaly. However, it has to have an anomaly as without it we have a projector $\eta_{\cal C}=e-2\frac{e+h_1+h_2+h_3}4$ which would have led to a critical point with $B=\frac14$: which we do not observe: this is the new feature of having an anomaly. We conclude that from the local perspective we have the ${\mathbb Z}_2\times {\mathbb Z}_2$ group with a mixed 't~Hooft anomaly.

\ 

Without the 't~Hooft anomaly the group law is
\be
g^2=h^2=e\,,\qquad g\,h=h\,g\,.
\ee The group manifold is $\U(1)^3$ and a general unitary operator is then
\be
&&g(\theta^1,\theta^2,\theta^3)=e^{i(\theta^1 g+\theta^2 h+\theta^3 gh)}=\\
&&\;\;\;\; (\cos\theta^1+i\sin\theta^1\,g)
(\cos\theta^2+i\sin\theta^2\,h)
(\cos\theta^3+i\sin\theta^3\,gh)\,.\nonumber
\ee This leads to
\be\label{eq:Bnoanomaly}
B(\theta^1,\theta^2,\theta^3)= \left(\cos\theta^1\cos\theta^2\cos\theta^3\right)^2+\left(\sin\theta^1 \sin\theta^2\sin\theta^3\right)^2\,.
\ee This gives us maxima corresponding to the identity operator (up to phases) and many minima with $B=0$. However, let us focus on critical points with $0<B<1$. The critical points are located at loci satisfying
\be
&&\sin2\theta^1 \left(\left(\sin\theta^2\sin\theta^3\right)^2-\left(\cos\theta^2\cos\theta^3\right)^2\right)=0\,,\\
&&\sin2\theta^2 \left(\left(\sin\theta^1\sin\theta^3\right)^2-\left(\cos\theta^1\cos\theta^3\right)^2\right)=0\,,\nonumber\\
&&\sin2\theta^3\left( \left(\sin\theta^2\sin\theta^1\right)^2-\left(\cos\theta^2\cos\theta^1\right)^2\right)=0\,.\nonumber\\
\ee These have an interesting set of solutions with $\cos\theta^i=\pm\frac1{\sqrt{2}}$. For all such solutions $B=\frac14$. These are the only solutions with $0<B<1$.
All of these solutions are proportional to the gauging ${\mathbb Z}_2$, $\frac{e-g-h-gh}2$, up to redefinitions of the group elements by phases keeping the fusion rules intact, as expected.\footnote{As in the ${\mathbb Z}_2$ Tambara-Yamagami case the redefinitions by phases can also be  understood as critical points for $g-\frac{e+g+h+gh}2$, $h-\frac{e+g+h+gh}2$, and $gh-\frac{e+g+h+gh}2$. See Section \ref{sec:simpleproofs}.} 

If one had obtained the $B$ of \eqref{eq:Bnoanomaly}, studying the minima would have led one to three different operators each forming a ${\mathbb Z}_2$. Finding then the saddles above is consistent with the ${\mathbb Z}_2\times {\mathbb Z}_2$ group.

\ 

\subsection{$Rep(S_3)$}

Finally let us discuss one additional example, ${\cal C}=Rep(S_3)$. 
The fusion is given by
\be\label{RepS3fusion}
D^2=e+\eta+D\,,\qquad \eta^2=e\,,\qquad D\,\eta=\eta\, D=D\,.
\ee The quantum dimension of $D$ is $2$. The general unitary operators take the form,
\be
&&g(\theta,\phi)=e^{i\phi \eta}e^{i\theta D}=\frac{e}3\left(\frac12 e^{2i\theta}+e^{-i\theta}+\frac32\right)+\\
&&\;\;\;\;\;\;\;\;\;\;\;\;\;\frac\eta3\left(\frac12 e^{2i\theta}+e^{-i\theta}-\frac32\right)+\frac{D}3\left(e^{2i\theta}-e^{-i\theta}\right)\,,\nonumber 
\ee and the group manifold is $\U(1)^2$. Using this we get
\be
B(\theta,\phi)=\frac{7+2\cos3\theta+6\cos(\theta-2\phi)+3\cos2(\theta+\phi)}{18}\,.
\ee  Studying critical points we have minima with $B=0$ at $g=\eta$ (up to phases) and maxima with $B=1$ at $g=e$ (up to phases). We have two types of critical points with $0<B<1$. The first is at $B=\frac{4}9$
and corresponds to the gauging projector (up to phases),
\be
g=e-2\frac{2D+e+\eta}{2^2+1^2+1^2}\;\; \to \;\; B=\left(1-\frac2{2^2+1^2+1^2}\right)^2=\frac49\,.
\ee The second one has $B=\frac19$ and is given by (up to phases),
\be
g=\eta-2\frac{2D+e+\eta}{2^2+1^2+1^2}\;\; \to \;\; B=\left(0-\frac2{2^2+1^2+1^2}\right)^2=\frac19\,.
\ee The second type of a critical point is an additional one we find in this case. Notice that if one would obtain this function $B$ and would try to deduce the category from it one immediately would know that there are three simples; two with dimension one and one with dimension $d(D)$. Then if one wants to interpret the point with $B=\frac19$ as gauging projector one would write
\be
\frac19=\left(1-\frac2{1^2+1^2+d(D)^2}\right)^2\;\;\to \;\; d(D)=1\,.
\ee However, if $d(D)$ is equal to one it is invertible, meaning we should have found additional points with $B=0$. Thus we deduce that $B=\frac{4}9$  is the gauging projector leading to $d(D)=2$. Note that 
the extra critical point we have is related to the fact that the category has a ${\mathbb Z}_2$ subgroup, see Section \ref{sec:simpleproofs}.

\section{Summary and discussion}\label{sec:discussion}

In this paper we have reviewed and commented upon some of the simplest features of non-invertible symmetries. Our discussion was centered on the apparent tension between unitarity and locality and its relation to  non-invertible symmetries. 

Concretely, we have considered local QFTs assuming that they have a categorical structure of symmetries satisfying the {\it regularity} condition \eqref{eq:regularitycondition}.  For such theories we have discussed an alternative approach to symmetry focused on unitarity connecting directly with  Wigner's approach.\footnote{The relation between non-invertible symmetries and Wigner's theorem, and invertible symmetries more generally, was discussed in the literature from various perspectives. See {\it e.g.} \cite{Cordova:2024iti,Ortiz:2025psr}. Typically one insists on locality and these relations involve embedding the non-invertible system into a larger invertible one with additional constraints. In fact, as we have already mentioned, gauging a unitary global symmetry is exactly a procedure of this sort which can lead to non-invertible symmetries. See also \cite{Casini:2020rgj,Benini:2025lav,AliAhmad:2025bnd,Lamas:2025eay} for related works on entanglement asymmetry and non-invertible symmetries. }  We studied the space of unitary symmetry operators $g$ and introduced an observable, $B(g)$ defined in \eqref{eq:regularityconditionG}, which measures deviations from locality. We focused on theories with finite symmetries leading to finite dimensional Lie groups.   
We also explored some generic properties of $B(g)$ and illustrated them with examples.

The space spanned by $g$ (the group manifold ${\cal M}_G$ and its dimension) and the function $B(g)$ are determined by the symmetry structure of the theory (including how it acts on the Hilbert space, and some of the anomalies). The space of theories splits into universality classes parametrized by the group $G$, with the classification further refined by the function $B(g)$. This function is fairly restricted, and interestingly encodes the data of the fusion algebra of the topological symmetries. 
Deforming the theory, some symmetries might be broken and thus one can restrict to a submanifold of ${\cal M}_G$ in the IR. On the other hand, some symmetries might also emerge so this submanifold might be naturally embedded in a bigger one. The structure of $B(g)$ thus encodes some universal properties of a class of theories related by RG flows. 
Lattice theories do not enjoy this universality as we have discussed and it only emerges together with locality in the continuum/thermodynamic limit.

One application of our approach that we envision is as a tool to understand which symmetries of a UV description, {\it e.g.} a lattice one or the UV regularized Lagrangian, 
will be manifest as unitary symmetries in  the continuum limit and which will give rise to non-invertible ones. There are several important subtleties here.
The first  is that on the lattice the space-time symmetry might mix with global ones. For example, one can have mixed anomalies between lattice translations and global symmetries \cite{Cheng:2022sgb} or group laws that mix the two, see Appendix \ref{app:etaplusdefect}. In our discussion here we have implicitly assumed that we can cleanly disentangle the two types of symmetry, but this is not always the case. Note also that operators in $2d$ CFTs which are (time) translation invariant but not local, were discussed in the literature. See {\it e.g.} \cite{Ambrosino:2025myh,Ambrosino:2025pjj}. The second subtlety is that on the lattice all translation-invariant symmetry operators could contribute to $B(g)$. Then an interesting question/subtlety is whether for a given theory all of these come from categorical symmetries or not. Our statements about $B(g)$ only hold for the subspace of all unitary operators which can be written as a combination of simple operators from the local perspective.

Another application can be from the point of view of the S-matrix bootstrap.  Here, the main objective is to characterize the S-matrix of quantum systems that maps an in-Hilbert space to an out-Hilbert space by using broad principles such as unitarity, analyticity, causality, and locality.  Symmetries, here, are the set of operators that commute with S-matrix,  and their unitarity properties are naturally realized. {\it A-priori} there is no assumption of a bulk spacetime, and therefore such symmetries may or may not be locally acting.  The trace observable of the symmetry operator in this paper may provide a diagnostic of locality of such operators that commute with the S-matrix.  In particular we hope that it can inform whether there could exist symmetries that are inherently non-local, meaning they do not admit descriptions in terms of topological operators.

There are numerous avenues for possible generalizations. For example, one can consider higher-form symmetries in dimensions higher than two.\footnote{See for example \cite{Harlow:2025cqc} for a discussion of locality in such setups.} The function $B(g)$ can be defined in any number of dimensions but the physical properties deducible from it will depend on the dimensionality of space-time. It would also be interesting to understand similar constructions in lattices of higher dimensional QFTs, {\it e.g.} quiver theories, {\it \'{a} la} \cite{ Razamat:2025wri}.

Although we have briefly discussed the case of a theory with a symmetry realized projectively, it would be highly desirable to understand how (and which) anomalies (and their generalizations, {\it e.g.} the associators in the $2d$ fusion category case) in general are captured in the unitary approach. See for example \cite{Delmastro:2021xox} for a relevant work.

We have considered finite symmetries. The generalization to continuous symmetries in the categorical context is poorly understood. In our unitary approach, continuous symmetries would lead to infinite dimensional Lie groups. 
Thus the function $B(g)$ would be defined over an infinite-dimensional space. We would like to understand statements one can make in such setups.

 Finally, of course, it will be of great interest to understand direct physical implications which are easier to access using the unitary approach. For example, 
 violation of \eqref{eq:regularitycondition} for unitary symmetries can usually be interpreted as an indication of persistent order. However, from our discussion it might be also a sign that the underlying UV symmetry might in fact possess non-invertible symmetries. For example, even without taking $\beta\to 0$ limit one can wonder whether imbalance of charges, {\it e.g.} baryon charge asymmetry, might be due to the associated symmetry
 being implemented locally as a non-invertible symmetry in the deep UV. We leave these and similar questions for future investigations.

\ 

\ 

\ 

\

\noindent{\bf Acknowledgments}:~
We are grateful to Arash Arabi-Ardehali, Cristopher Beem, Behzat Ergun,  Minjae Kim, David Jordan, Zohar Komargodski, Konstantinos Roumpedakis, Sakura Sch\"{a}fer-Nameki, Shu-Heng Shao, Amos Yarom, Thomas Waddleton, Yifan Wang, Wenbin Yan, Gabi Zafrir and  Yunqin Zheng for insightful discussions and comments.  IB and SSR are particularly grateful to Zohar Komargodski for hosting us at SCGP, and for watching and entertaining our children during the course of this project!
 The research of SSR and MS is supported in part by  the Planning and Budgeting committee, by the Israel Science Foundation under grant no. 2159/22, and by BSF-NSF grant no. 2023769.  IB, HT are supported
in part by the Simons Collaboration on Global Categorical Symmetries and also by the NSF grant PHY-
2412361.

\appendix

\section{Some comments on the transverse field Ising model }\label{app:etaplusdefect}
Consider the  Hamiltonian of the transverse field Ising model defined on a periodic chain,
\begin{align}
    H = -\sum_{s=1}^{L}\left(X_{s} + Z_{s}Z_{s+1}\right),\qquad Z_{L+1} = Z_{1},
\end{align}
where we have \(L\) lattice sites and $X$ and $Z$ are Pauli matrices. We choose the number of lattice points to be odd for later convenience; however the results can be adjusted for an even number of sites.  Here we will mainly follow the conventions and results derived in \cite{Cheng:2022sgb,Seiberg:2023cdc,Seiberg:2024gek}. This model possesses several symmetries. First, it has translational invariance, implemented by \(T\),
\begin{align}
    T: \mathcal{O}_{i} \rightarrow \mathcal{O}_{i+1},\qquad \mathcal{O} = \{X,Z\},
\end{align}
for \(i=1,\dots,L\) with periodic boundary conditions, \textit{i.e.} \(\mathcal{O}_{L+1} = \mathcal{O}_{1} \). It is implemented by the following unitary operator which satisfies a \(\mathbb{Z}_{L} \) group-law,
\begin{align}
    T = \prod_{i=1}^{L-1} S_{i,i+1},\qquad S_{i,j} = \frac{1}{2}\left(1+X_{i}X_{j} + Y_{i}Y_{j} + Z_{i}Z_{j}\right),\qquad T^{L} = 1.
\end{align}
The model also has additional symmetries: one is a group-like  symmetry, \(\eta\),
\begin{align}
    \begin{aligned}
        \eta: Z_{i}\rightarrow -Z_{i}\, \qquad \eta: X_{i} \rightarrow X_{i},
    \end{aligned}
\end{align}
for all \(i=1,\dots,L \),  implemented by the following unitary (and hermitian) operator which satisfies a \(\mathbb{Z}_{2} \) group-law,
\begin{align}
    \eta = \prod_{i=1}^{L} X_{i},\qquad \eta^{2} = 1.
\end{align}
Another is a non-invertible symmetry,  denoted by \(D\), which exchanges the \(X\) and \(ZZ\) terms in the Hamiltonian,
\begin{align}
    \begin{aligned}
        D&: X_{i} \rightarrow Z_{i}Z_{i+1},\\
        D&:Z_{i-1}Z_{i} \rightarrow X_{i},
    \end{aligned}
\end{align}
for \(i=1,\dots,L\). In contrast with the previous symmetries, this one is implemented by a non-invertible operator,
\begin{align}
\begin{aligned}
        D &=  U_{KW} \frac{1+\eta}{2}\,, \qquad U_{KW} &= \text{e}^{-\frac{i \pi L}{4}}\left( \prod_{s=1}^{L-1}\frac{1+i X_{s}}{\sqrt{2}} \frac{1+ i Z_{s}Z_{s+1}}{\sqrt{2}} \right) \frac{1+ i X_{L}}{\sqrt{2}}\,,
\end{aligned}
\end{align}
where \(\frac{1+\eta}{2} \) is a projector making the operator non-invertible, while  \(U_{KW}\) is unitary. 

\ 

\noindent These basic symmetry operators have the simple multiplication rules
\begin{align}
\begin{aligned}
        &T^{L} = 1,\qquad &&\eta^{2} = 1,&&&\qquad D^{2} = \frac{1+\eta}{2} T,&&&&\qquad D^\dagger=D T^{-1}\,,\\
    &T\eta = \eta T, &&TD=DT, &&&D\eta = \eta D =D\,,&&&&
    T^\dagger=T^{-1}\,.
\end{aligned}
\end{align}
The operators  \(T\) and \(\eta\) are also unitary operators while \(D\) is not.
Note that \(T,\eta\) and \(D\)  map local operators that appear in \(H\) to local operators. However, when acting on $Z_j$ the operator $D$ maps it to a string of operators,
\begin{align}
    D\,Z_{j}= \left(U_{KW}\,\frac{1+\eta}{2}\right)\,Z_{j} = \left(i X_{j}X_{j-1}\cdots X_{1}Z_{1}\right)\, \left(U_{KW}\, \frac{1-\eta}{2}\right).
\end{align}
When \(D\) acts on \(Z_{j}\) it flips the projection in \(D\) since \(Z_{j}\) is negatively charged under \(\eta\). In the thermodynamic limit $L\to \infty$ this system becomes the continuum Ising model CFT with $T$ mapping to infinitesimal translation while $D$ maps (up to normalization) to the KW non-invertible defect.  Note that the fusion algebra is slightly different to the one appearing in the bulk of the paper. The hermiticity of $D$ fails by a factor of $T$ and $D^2$ is different from the continuum one also by a factor of $T$. As $T$ becomes 
infinitesimal translation in the continuum limit the hermiticity and the algebra emerge. However, this is an important example where on the lattice the space-time symmetries and global symmetries are not cleanly disentangled.

\

\noindent One can use \(D,\eta \) and \(T\) to build general symmetries on the lattice. As the space-time and global symmetries are entangled here this exercise is quite interesting. For example, let us try to construct a unitary operator on the lattice which would become $\eta_+$ in the continuum. Up to normalization a natural candidate is
\begin{align}
    \eta_{+} = D + \frac{1-\eta}{2}\,.
\end{align} 
However, notice that this operator in fact generates a ${\mathbb Z}_{2L}$ symmetry and not a ${\mathbb Z}_2$ because the fusion of two $D$s involves a $T$. In particular we can define
\be
\eta^\pm_k =D\, T^k\pm \frac{1-\eta}2\,,\qquad e^\pm_k =\frac{1+\eta}2\, T^k\pm\frac{1-\eta}2\,\;\;\; (k=0,\dots,L-1)\,,
\ee which satisfy a simple ${\mathbb Z}_{2L}\times {\mathbb Z}_2$ group law. For example,
\be
\eta^+_k\eta^+_l=e^+_{k+l+1}\,,\qquad 
\eta^+_k e^+_l=\eta^+_{k+l}\,,\qquad
e^+_ke^+_l=e^+_{k+l}\,.
\ee Here $\eta^+_0$ generates the ${\mathbb Z}_{2L}$ group while $e^-_0=\eta$ generates the ${\mathbb Z}_2$.\footnote{Note that we have  $(1-\eta^+_0)(1-\eta)=0$.} Thus we see that we cannot really disentangle the group manifold of the global symmetries and the space translations on the lattice at any finite $L$. Note that there is also an interesting difference between odd and even $L$. In the even case, $L=2n$, the natural ${\mathbb Z}_2$ subgroup of ${\mathbb Z}_{2L}$ is generated by $e^+_n$.
For odd $L=2m+1$, ${\mathbb Z}_{2L}={\mathbb Z}_2\times {\mathbb Z}_L$. The ${\mathbb Z}_L$ is generated by $e^+_1$ while the ${\mathbb Z}_2$ subgroup is generated by
\be
\hat \eta_\pm\equiv \eta^\pm_m=D\, T^m\pm\frac{1-\eta}2\,,\qquad 
\left(\hat\eta_+\right)^2=1\,,\qquad
\hat\eta_+\,\hat\eta_-=\eta\,.
\ee 
Thus at least in the $L$ odd case there is a very natural group manifold which is similar in many respects to the continuum one generated by
${\mathbb Z}_2\times {\mathbb Z}_2$ elements $\hat\eta_+$ and $\eta$. This is a $\U(1)^2$ group manifold which becomes  isomorphic to the continuum one in the $m\to\infty$ limit. More precisely, following \cite{Seiberg:2024gek},
\be\label{eq:contlatmap}
\eta_{latt.}\to \eta_{cont.}\,,\qquad D_{latt.}\to \frac{1}{\sqrt{2}} \, D_{cont.}\,e^{\frac{2\pi i\, P}{2L}}\,,\qquad T\to e^{\frac{2\pi i\, P}{L}} \,, \qquad P=h-\bar h\,,
\ee with $P$ being the generator of continuum translations. From here we thus write  that
\be
\hat \eta_\pm  \to \frac1{\sqrt{2}}\,D_{cont.} e^{\frac{2\pi i\, P(2m+1)}{2L}}\pm \frac{1-\eta_{cont.}}2
=
\frac{(-1)^{h-\bar h}}{\sqrt{2}}\,D_{cont.}\pm \frac{1-\eta_{cont.}}2\,.
\ee The operator $\hat \eta_\pm $ differs then from the $\eta_\pm$ we discussed in the bulk of the paper by the factor of $(-1)^{h-\bar h}$.
Let us next study some properties of $\hat \eta_\pm$.

\

We concentrate on $\hat \eta_+$ without loss of generality.
This operator does not send local operators to local operators. For example, we consider  transformations of local operators $X_i$, which correspond to sites, and $Z_i Z_{i+1}$, which correspond  to links \cite{Seiberg:2023cdc}, 
\be \label{eq:epact}
\hat\eta_+\, X_j \hat\eta_+ &&=Z_{j+m+1}Z_{j+m}\,\frac{1+\eta}2+X_j\,\frac{1-\eta}2\,, \\
\hat\eta_+\, Z_j Z_{j+1} \hat\eta_+ &&=X_{j+m+1}\,\frac{1+\eta}2+Z_j Z_{j+1} \,\frac{1-\eta}2\,.\nonumber
\ee  The non-locality is  manifest in the fact that the action of $\hat\eta_+$ on local operators generates a superposition of operators fused with topological operators, in this case $\eta$.  This is an important feature of a non-invertible symmetry, {\it i.e.} it mixes the kinematics of local operators with non-local ones, thus imposing constraints between the spectrum of the two types of objects.  In the context of representation theory of categorical symmetries such a mixing of local and non-local operators was recently studied and discussed in several contexts, see {\it e.g.} \cite{Cordova:2024iti,Ueda:2025ecm}.  The $\hat\eta_+$ symmetry operator makes this point explicit and transparent from the point of view of the unitary approach.

One can study the eigenstates of $H$ and their quantum numbers under $\hat\eta_+$ and $\eta$. For example, one can compute the partition function of the theory,
\be
Z(\beta,m,n, k, L)=\Tr\, e^{-\beta\, H}\, {\hat\eta}_+^{m} \,\eta^n\, T^k\,.
\ee The lattice theory is a vanilla quantum mechanical system and this partition function is completely well-defined. In particular, for $(\mathfrak{E}_+,\mathfrak{O}_+)$ equal to number of even $\hat\eta_+$ and odd $\hat\eta_+$ states respectively, we can evaluate the ratio,
\be 
B(\hat\eta_+;L)=\left|\frac{Z(0,1,0,0,L)}{Z(0,0,0,0,L)}\right|^2=\left|\frac{\mathfrak{E}_+-\mathfrak{O}_+}{\mathfrak{E}_++\mathfrak{O}_+}\right|^2\,,
\ee
and show that it approaches $1/4$ for large values of $L$; that is, $\mathfrak{E}_+$ is $3$ times $\mathfrak{O}_+$. For finite values of $L$ this ratio is different from $1/4$ but approaches it as we increase $L$. Computing this explicitly for low $L$ we obtain results consistent with $B(\hat\eta_+;L=2m+1)=\left(\frac{2^{m}+1 }{2^{m+1}}\right)^2$.  A qualitative way to understand this is as follows. The ratio of $\eta$ even and odd states is approximately $1:1$. However, 
because of the fact that $(1-\eta)(1-\hat\eta_+)=0$, all $\eta$ odd states are $\hat\eta_+$ even.
The $\eta$ even states however can be either $\hat\eta_+$ odd or even. We can compute also the function $B$ as a function of $m$ and see how it approaches the continuum. First, because $\sqrt{2}DT^m$ and $\eta$ satisfy exactly the same fusion algebra as ${\mathbb Z}_2$ Tambara-Yamagami,
\be
\left(DT^m\right)^\dagger=DT^m\,,\;
\left(\sqrt{2}DT^m\right)^2=1+\eta\,,\; \eta \left(DT^m\right)=\left(DT^m\right)\eta=\left(DT^m\right)\,,\; \eta^2=1\,,\;\;\;\;
\ee the general group element built from this is the same as $g(\theta,\phi)$ of
\eqref{eq:TYZ2group},
\be \label{eq:TYZ2groupLattice}
&&g_m(\theta,\phi)=e^{i\theta\,\eta}\,e^{i\phi\, \sqrt{2}DT^m}=\\
&&\;\;\;\;\frac12\left(\cos\sqrt{2}\phi \,e^{i\theta} +e^{-i\theta}\right)\,e+
\frac12\left(\cos\sqrt{2}\phi \,e^{i\theta}-e^{-i\theta}\right)\,\eta+i\sin\sqrt{2}\phi e^{i\theta}\, DT^m\,.\nonumber
\ee  The group manifold is thus again $\U(1)^2$. However,
\be
&&B_m(\theta,\phi)=\\&&\;\;\frac{1+\cos^2\sqrt{2}\phi+2\cos2\theta \cos\sqrt{2}\phi}4-\left(\frac{\Tr\,DT^m}{2^{2m+1}}\right)\sin\sqrt{2}\phi\sin2\theta+
\left(\frac{\Tr\,DT^m}{2^{2m+1}}\right)^2\sin^2\sqrt{2}\phi\,.\nonumber
\ee Here we already took into account that $\Tr\,\eta=0$ also on the finite dimensional lattice. From the result that $B(\hat\eta_+;L=2m+1)=\left(\frac{2^{m}+1 }{2^{m+1}}\right)^2$ we deduce
\be
\left(\frac{\Tr\,DT^m}{2^{2m+1}}\right)=
\frac{2^{m}+1 }{2^{m+1}}-\frac12=\frac1{2^{m+1}}\,,
\ee and thus,
\be 
B_m(\theta,\phi)=\frac{1+\cos^2\sqrt{2}\phi+2\cos2\theta \cos\sqrt{2}\phi}4-\frac{\sin\sqrt{2}\phi\sin2\theta}{2^{m+1}}+
\frac{\sin^2\sqrt{2}\phi}{2^{2m+2}}\,.\;\;
\ee We see that the lattice result approaches exponentially the continuum value of $B$.

\

\

%\newpage

\bibliographystyle{jhep}

\bibliography{refs}

\end{document}